\documentclass[12pt]{article}
\usepackage{amsfonts}
\usepackage{amsmath}
\usepackage{graphicx}
\usepackage{multirow}
\usepackage{mathrsfs}
\usepackage{color}
\newcommand{\ud}{\mathrm{d}}
\newcommand{\ui}{\mathrm{i}}

\newcommand{\Real}{\mathrm{Re}}
\newcommand{\Imag}{\mathrm{Im}}
\newcommand{\expp}{\mathrm{e}}

\newcommand{\corresponds}{\mathrel{\widehat{=}}}

\begin{document}
\title{Trace formula for a dielectric microdisk  with a point scatterer}

\date{} % delete this line to display the current date

\author{Robert F. M. Hales$^1$\thanks{Robert.Hales@bristol.ac.uk}, Martin Sieber$^1$ and Holger Waalkens$^{2}$}

\maketitle
\noindent
{\small $^1$ School of Mathematics, University of Bristol, University Walk, Bristol BS8 1TW, UK}\\
{\small $^2$ Johann Bernoulli Institute for  Mathematics and Computer Science, University of Groningen, 
PO Box 407, 9700 AK Groningen, The Netherlands} \bigskip

\begin{abstract}
%\new{
Two-dimensional dielectric microcavities are of widespread use in microoptics applications. Recently, a trace
formula has been established for dielectric cavities which relates their resonance spectrum to the periodic rays
inside the cavity. In the present paper we extend this trace formula to a dielectric disk with a small scatterer.
This system has been introduced for microlaser applications, because it has long-lived resonances
with strongly directional far field. We show that its resonance spectrum contains signatures not only of periodic
rays, but also of diffractive rays that occur in Keller's geometrical theory of diffraction. We compare our results with 
those for a closed cavity with Dirichlet boundary conditions.

\end{abstract}
\bigskip


\section{Introduction}\label{sec:intro}

Two-dimensional optical microcavities are flat dielectric objects that are surrounded by a medium of lower
refractive index. They trap light by total internal reflection and are of great interest as miniature lasing devices 
with many technological applications \cite{Vahala2004}. The confinement of light results in long-lived states
which is key for achieving low lasing threshold. In addition to the light confinement another highly desirable
design feature is the directionality of the output.
 Examples of recent studies to achieve these goals include deformations of the boundary
\cite{Dubertrand2008, Shinohara2009} or insertion of a hole \cite{Wiersig2006} or a defect \cite{Apalkov2004,
Tulek2007,Dettmann2008}. A more detailed list of references can be found in these articles.

The emission patterns of dielectric cavities are determined by the eigenmodes that are selected during the pumping
process. Because dielectric cavities are open systems, the eigenmodes are resonance states with complex 
wavenumbers $k=k_r+\ui k_i$. The real part of the wavenumber is related to the frequency $\omega$ by $k_r=\omega/c$,
where $c$ is the speed of light, and the imaginary part is related to the lifetime $\tau$ by $k_i=-1/2 c \tau$.
In practice, the resonance spectrum of a cavity has to be determined numerically or experimentally,
because the only analytically solvable systems are the integrable cases with rotational symmetry like the circle. One interesting
question is how one can understand the distribution of resonances in the complex plane that are found by
the numerical or experimental methods. For this purpose, it has been very useful to adapt short-wavelength
approximations that are of widespread use in the field of quantum or wave chaos \cite{Stoeckmann1999}.
In a recent paper short-wave methods have been applied to determine a Weyl law for the number of
resonances, a lower bound for their imaginary part, and to reveal signatures of periodic
rays in the resonance spectrum \cite{Bogomolny2008}.

In this article we will focus on the trace formula for dielectric cavities that was obtained in \cite{Bogomolny2008}.
It considers the excess density of states which is a sum over Lorentzian terms for the resonances. In the
short-wave approximation this quantity is approximated by a smooth Weyl term, describing the mean
density of resonances, plus a sum over the periodic rays inside the cavity. So far its main application
has been to consider Fourier transforms of resonance spectra and identify the positions of its peaks
with lengths of periodic orbits \cite{Bittner2010,Bogomolny2010}.

Trace formulas for dielectric cavities are on a less firm theoretical footing than the related
trace formulas for closed systems that were developed in the 1970's by Gutzwiller and Balian/Bloch
\cite{Balian1970,Balian1971,Balian1972,Gutz1971,Gutz1990}. There is not yet a systematic way
of deriving it, except for the integrable circular disk. It is known to require corrections for periodic
rays with reflection angles that are close to the critical angle of total reflection. This leads to large
finite size effects if only a small part of the resonance spectrum is known. A test of the true short-wave
regime requires a large number of resonances which are difficult to obtain except for the circular disk.

\begin{figure}
\centerline{
\includegraphics[angle=0,width=5cm]{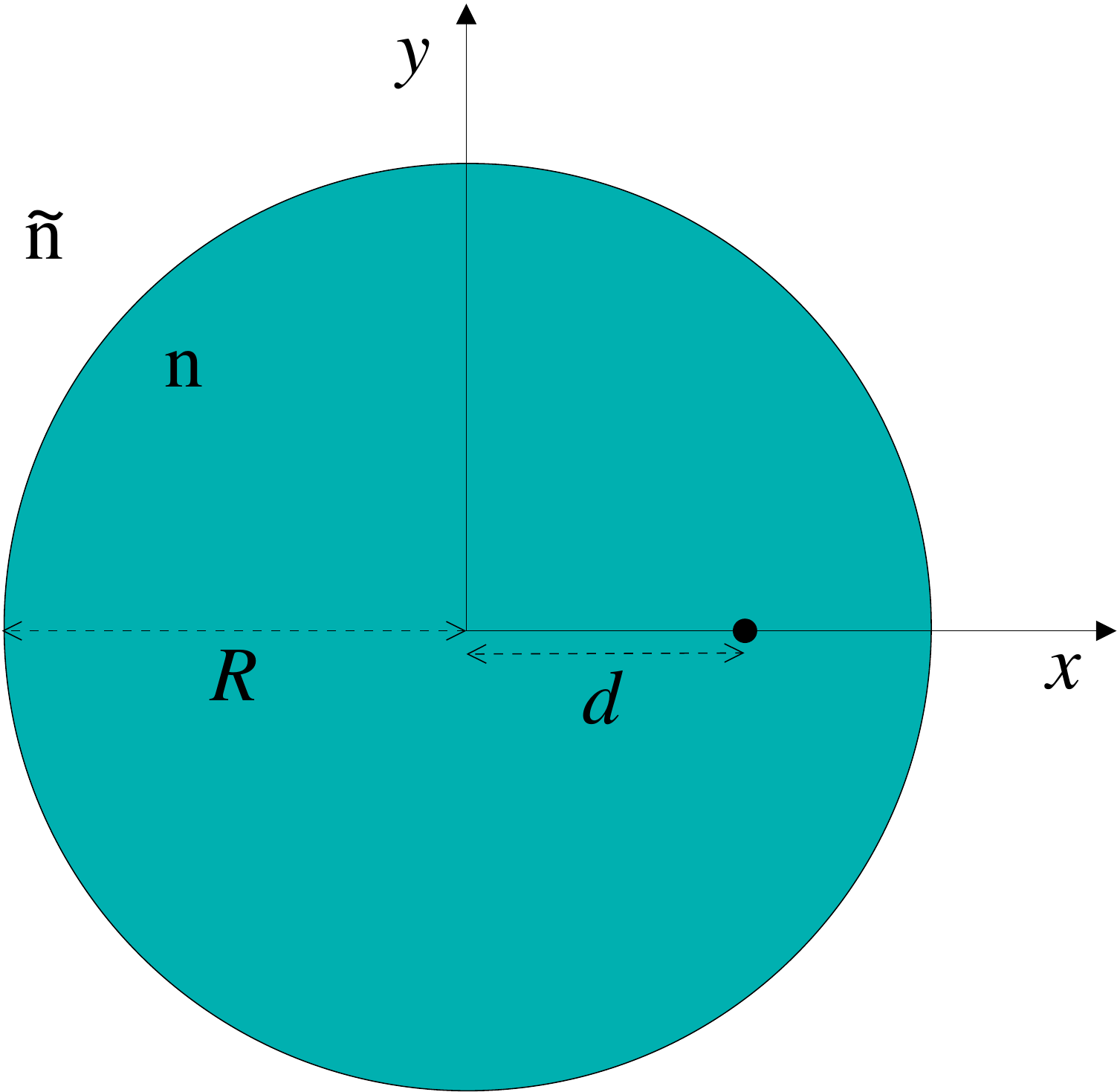}
}
\caption[Circular microcavity of radius $R$ and refractive index $n$.]{\label{fig:circle_cavity}
Circular dielectric cavity of radius $R$ and refractive index $n$ in an outer medium of refractive index $\tilde{n}$.
For the closed system we impose Dirichlet boundary conditions at the interface. For the perturbed systems a
point scatterer is added on the $x$-axis at a distance $d$ from the centre of the cavity. Throughout this 
article we will take $\tilde{n}=1$.
}
\end{figure}

In this article we will study the trace formula for a dielectric circular disk with a point scatterer, see 
Fig.~\ref{fig:circle_cavity}. It was shown in \cite{Dettmann2008,Dettmann2009uni} that this system
has resonance modes with highly directional far field in addition to large quality factors
$Q=-k_r/2 k_i$. For our study it has the advantage that it is non-integrable but allows,
nevertheless, the determination of a large number of resonances by a Green function method.

The addition of a point scatterer leads to additional contributions in the trace formula from 
diffractive rays which are closed orbits that begin and end at the scatterer. These rays appear in
Keller's geometrical theory of diffraction \cite{Keller1962}. They have been studied in detail for closed billiards
\cite{VatWirRos1994,Pavloff1995, Sieber1999, RahavRichFish2003, RahavFish2002,Laurent2006},
and we adopt these results here for open dielectric systems. The closed orbits in a circular cavity 
have been recently classified in \cite{Brack2009}. They undergo bifurcations as the position
of the scatterer is changed which we will treat by uniform approximations. We will also 
consider the dielectric circle without scatterer, and we compare our results to those for
closed cavities with Dirichlet boundary conditions (mirror boundary).

Our article is organized as follows. After an introduction of the trace formula in section ~\ref{sec:trace}
we first consider the closed cavity without scatterer in section~\ref{sec:cavity_closed},
and with scatterer in section~\ref{sec:ptbd_closed}. We then investigate the open dielectric
cavity without scatterer in section~\ref{sec:unptbd_open}, and with scatterer in section~\ref{sec:ptbd_open}. 
Section~\ref{sec:conc} contains our conclusions.

\section{Trace formulas: from closed to open cavitities}\label{sec:trace}

A common approach for studying the emission properties of microresonators is to consider
them as passive systems and neglect the interaction between the electromagnetic field
and the cavity material. The electromagnetic field for these dielectric cavities
is then described by the three-dimensional vectorial Maxwell equations.

Maxwell's equations simplify considerably if the electromagnetic field does not depend
on one spacial direction, say $z$, in systems with cylindrical symmetry \cite{Jackson1998}.
The electromagnetic field can then be decomposed into two independent polarizations,
called TM and TE modes. For each of these polarizations, Maxwell's equations can be
reduced to a two-dimensional scalar equation in the $xy$-plane. For homogeneous
dielectric cavities with refractive index $n$ and cross-section $D$, this equation has the form,
see e.g.\  \cite{Dettmann2009uni},
\begin{equation}\label{eq:hholtz_2d}
\left( \nabla^2 + n^2(\mathbf{r}) \, k^2 \right) \, \psi(\mathbf{r}) = 0 \, , \qquad
n(\mathbf{r}) = \begin{cases}
n,  \quad & \mathbf{r} \in D, \\
1,  \quad & \mathbf{r} \notin D,
\end{cases}
\end{equation}
where $\mathbf{r}=(x,y)$ and $k$ is the wavenumber. Here it is assumed that the surrounding medium has refractive index 1 (like air).
The function $\psi(\mathbf{r})$ describes the $z$-component of the
electric field  in the case of TM modes, and the $z$-component of the magnetic field  in the
case of TE modes. All other components of the electromagnetic field can be obtained from
the solutions of (\ref{eq:hholtz_2d}). The two modes satisfy different connection conditions
at the boundary of the cavity. In this article we will only consider TM modes for which
$\psi(\mathbf{r})$ and its derivatives are continuous across the boundary of the domain $D$.

The reduction of Maxwell's equations to equation (\ref{eq:hholtz_2d}) requires cylindrical 
symmetry, but it is also approximately valid in thin dielectric cavities in which the 
height is of the order of the wavelength. In this case $n$ is an effective refractive
index that depends on the material as well as the thickness of the cavity. Although
theoretical models exist for this index \cite{Lebental2007}, it is usually determined
experimentally.
Errors of this two-dimensional approximation have been investigated in \cite{Bittner09}.

Equation (\ref{eq:hholtz_2d}) is the basic equation that is widely used for the
study of dielectric cavities. It describes open systems in which the wavefunctions
extend to infinity. A related closed system is given by the Helmholtz equation for
waves in a bounded domain $D$
\begin{equation} \label{eq:hholtz_qm}
\left( \nabla^2 + k^2 \right) \, \psi(\mathbf{r}) = 0 \, , \qquad \mathbf{r} \in D,
\end{equation}
with Dirichlet boundary conditions for which the wavefunction vanishes at the
boundary. This models, for example, an optical cavity with refractive index
$n=1$ and a mirror boundary. Alternatively, it also agrees with
the time-independent Schr\"odinger equation for {\em quantum billiards}.
These are quantum versions of systems in which a point particle moves freely
inside the domain $D$ and is specularly reflected at the boundary. 
Quantum billiards have been studied extensively, and asymptotic methods for
the semiclassical (short-wavelength) limit $k \rightarrow \infty$ are well established.
These methods have been very helpful in obtaining related short-wavelength
approximations for dielectric cavities. For this reason we will discuss trace formulas
for quantum billiards before we return to open dielectric cavities.

The density of states is a useful characterisation of the eigenvalue spectrum
of the Helmholtz equation and it is given by 
\begin{equation} \label{eq:dens_states_main}
d(k) = \sum_{m=1}^\infty \delta(k - k_m) \, , \quad k>0 ,
\end{equation}
where $k_m$ are the positive $k$-values for which equation (\ref{eq:hholtz_qm}) with
Dirichlet boundary conditions has non-trivial solutions. The density of states is 
formally related to the Green function of the system by 
\begin{equation}\label{eq:dos_green}
d(k) = -\frac{2 \, k}{\pi} \; \Imag \phantom{~} \mathrm{tr} \, G(\mathbf{r},\mathbf{r}',k) \, ,
\end{equation}
where the Green function is the solution of
\begin{equation} \label{eq:hholtz_green}
\left( \nabla^2 + k^2 \right) \, G(\mathbf{r},\mathbf{r}',k) = \delta(\mathbf{r} - \mathbf{r'}) \, ,\qquad 
\mathbf{r}, \mathbf{r'} \in D,
\end{equation}
with Dirichlet boundary conditions. In the limit where the wavelength is short compared
to the dimensions of the billiard the Green function can be approximated by a sum
over all classical trajectories connecting $\mathbf{r'}$ to $\mathbf{r}$ in the billiard system.
If one then evaluates the trace in (\ref{eq:dos_green}) in the asymptotic regime of
large wavenumbers one finds that the density of states is approximated by a
sum of a smooth part $d_0(k)$ and an oscillatory part $d_{\mathrm{osc}}(k)$
\cite{Balian1970,Balian1971,Balian1972,Gutz1971,Gutz1990}
\begin{equation}\label{eq:trace_formula}
d(k) \approx d_0(k)+d_{\mathrm{osc}}(k) \,.
\end{equation}
The smooth part comes from the contributions of orbits of zero length. It is given
in terms of the area $\mathcal{A}$ and perimeter $L$ of the billiard system by Weyl's law \cite{BH76}
\begin{equation}\label{eq:trace_smooth_closed}
d_0(k)=\frac{\mathcal{A} k}{2 \pi} - \frac{L}{4 \pi} + o(1) \, .
\end{equation}
The oscillatory part consists of a sum over the classical periodic orbits in the billiard system.
In this article we will consider also perturbations of quantum billiards and dielectric cavities
by a point-scatterer inside the domain $D$. In this case there are additional contributions
to the oscillatory part from so-called diffractive orbits. These are trajectories which start and
end at the scatterer. Between start and end point they can have multiple reencounters
with the scatterer, and at each encounter the outgoing and incoming angles are
unrelated to each other. This can be related to the fact that an $s$-wave scatterer scatters
uniformly in all directions. We hence write the oscillatory part as
\begin{equation}\label{eq:trace_oscpart}
d_{\mathrm{osc}}(k)=d_{\mathrm{po}}(k)+d_{\mathrm{do}}(k)\,,
\end{equation}
where $d_{\mathrm{po}}(k)$ gives the contribution of periodic orbits and
$d_{\mathrm{do}}(k)$ gives the contribution of diffractive orbits.
For the unperturbed cases (without scatterer) we have that $d_{\mathrm{do}}(k)=0$.
Both periodic and diffractive trajectories give oscillatory terms to the density of
states of the form
\begin{equation} \label{eq:d_xi}
d_\xi(k) = \chi_\xi \, A_\xi(k) \,\expp^{\ui \, k \, l_\xi} + \mathrm{c}.\mathrm{c}\,,
\end{equation}
where $l_\xi$ is the length of the trajectory $\xi$, $A_\xi$ is an amplitude factor, and
$\mathrm{c}.\mathrm{c}.$
denotes the complex conjugate. Furthermore,
$\chi_\xi$ is a phase factor that arises from the Dirichlet boundary conditions. It
 equals $(-1)^{r_\xi}$ where $r_\xi$ is the total number of reflections at
the boundary. The contribution of a diffractive trajectory is weaker than that of a periodic one.
The amplitude contains a factor $k^{-1/2}$ for every scattering at the scatterer.

In \cite{Bogomolny2008} it was investigated how one can generalize the trace
formula to dielectric cavities. This is not obvious, because dielectric cavities are
open systems and equation (\ref{eq:hholtz_2d}) admits a continuous spectrum
corresponding to the scattering solutions. Using the Krein formula
\cite{Krein53,Krein62} it was shown that the relevant quantities to consider are
the resonance states. These are solutions of equation (\ref{eq:hholtz_2d})
with outgoing boundary conditions
\begin{equation}
\psi(\mathbf{r}) \; \propto \; \expp^{\ui  k |\mathbf{r}|} \, , \qquad |\mathbf{r}| \rightarrow \infty \, .
\end{equation}
They exist only for a discrete set of complex values of $k$ with negative imaginary part.
These complex $k$-values coincide with the poles of the scattering matrix, and also
with the poles of the Green function, and we denote them again by $k_m$.
It was shown in \cite{Bogomolny2008} that  for dielectric cavities,
one can again obtain a trace formula of the form (\ref{eq:trace_formula}) if one replaces the
delta-functions on the left-hand side by Lorentzians
\begin{equation}\label{eq:dstate_res}
d(k) = -\frac{1}{\pi} \sum_{k_m} \frac{\Imag k_m}{(k - \Real k_m)^2 + (\Imag k_m)^2} \,.
\end{equation}
There is, however, an issue concerning which resonances to include in this sum.
For TM modes one can typically distinguish two kind of resonances whose
wavenumbers are well separated in the complex plane. The inner or Feshbach
resonances have a high concentration inside the cavity and their wavenumbers
lie in a strip $-\gamma < \Imag k_m < 0$ where $\gamma$ depends on the
cavity. The outer or shape resonances are mainly concentrated outside the 
cavity and their wavenumbers have larger negative imaginary parts. Hence
the outer resonances give only a smooth contribution to the density of states.
These two types of resonances differ also in their behaviour as the refractive
index $n \rightarrow \infty$ as will be discussed later.

In \cite{Bogomolny2008} the smooth density of states $d_0(k)$ on the right-hand
side of the trace formula (\ref{eq:trace_formula}) was derived for the inner
resonances. Hence one should also include in the sum (\ref{eq:dstate_res})
only the inner resonances. The result of \cite{Bogomolny2008} for $d_0(k)$ was 
inferred from the example of a circular dielectric cavity which is integrable
and is given by
\begin{equation} \label{eq:trace_smooth}
d_{0}(k) = \frac{n^2 \mathcal{A} k}{2 \pi} + \tilde{r}(n) \frac{L}{4 \pi} \, ,
\end{equation}
where the function $\tilde{r}(n)$ has the form
\begin{equation} \label{eq:r_tilde}
\tilde{r}(n) = 1 + \frac{n^2}{\pi} \int_{-\infty}^{\infty} \frac{R(t)}{t^2+n^2} \ud t
- \frac{1}{\pi} \int_{-\infty}^{\infty} \frac{R(t)}{t^2+1} \ud t 
\end{equation}
with
\begin{equation}\label{eq:r_tilde_coeff}
R(t) = \frac{\sqrt{t^2+n^2} - \sqrt{t^2+1}}{\sqrt{t^2+n^2} + \sqrt{t^2+1}} \,.
\end{equation}
The oscillatory part in the trace formula is determined by the ray dynamics
inside the cavity. Since the contributing objects are formally the same as the
trajectories in billiards we will adopt the same terminology and speak of trajectories
or orbits instead of rays. In this language, the contributing trajectories are
exactly the same as in the closed cavity. The contribution of any single
trajectory is modified from (\ref{eq:d_xi}) to
\begin{equation} \label{eq:d_xi2}
d_\xi(k) = n \, R_\xi \, A_\xi(n k) \, \expp^{\ui \, n \, k \, l_\xi} + \mathrm{c}.\mathrm{c}. \, .
\end{equation}
Apart from the additional factor given by the refractive index $n$, the main difference
to the closed case
is that the phase factor $\chi_\xi$ is replaced by the quantity $R_\xi$ which
is the product of the Fresnel reflection coefficients for all the reflections 
of the trajectory at the boundary. If periodic orbits come in continuous
families, then one should replace $R_\xi$ by an average $\langle R_\xi \rangle$
over the family. For comparison with  \cite{Bogomolny2008}, note that we
consider $d(k)$ instead of $d(E) = d(k)/2 k$. 

The trace formula for dielectric cavities is on a less firm basis than that
for quantum billiards. The trajectory contributions (\ref{eq:d_xi}) have been
derived only for the case of the integrable circular cavity, and they have
been generalized to other cavity shapes by physical considerations. It is
also known that the contribution (\ref{eq:d_xi}) becomes inaccurate if one
of the reflection angles of a trajectory is close to the critical angle of total
reflection \cite{Jackson1998}
\begin{equation}\label{eq:crit_angle}
\theta_c(n)=\sin^{-1}\left(\frac{1}{n}\right) \,,
\end{equation}
where the reflection angle is defined with respect to the normal to the
boundary. This inaccuracy is due to the Goos-H\"anchen effect and
Fresnel filtering \cite{Bogomolny2008}. 

For TE modes, the separation of resonances into inner and outer
resonances is less obvious than for TM modes
\cite{Dubertrand2008,Dettmann2008},
and a trace formula has not yet been derived for TE modes.

We note that for all our numerical computations in this paper we choose the radius $R$ of the cavity to be 1. 
This is no restriction as the Equations~(\ref{eq:hholtz_2d}) and (\ref{eq:hholtz_qm}) (together with their boundary conditions) scale with $R$.

\section{The closed circular cavity} \label{sec:cavity_closed}

Short wavelength approximations are much better established for closed cavities than for open dielectric
cavities. We consider in the next two sections a circular disk with Dirichlet boundary conditions,
first without and then with perturbation. This allows us to introduce asymptotic short-wave approximations
in a well-established setting and it will illustrate the accuracy of the approximations in these cases.
It will set a benchmark for comparison with results for the dielectric disk discussed in Sections~\ref{sec:unptbd_open} and \ref{sec:ptbd_open}.

The Helmholtz equation (\ref{eq:hholtz_qm}) for a circular disk with radius R is separable in polar
coordinates, and we label the (unnormalized) solutions as
\begin{equation}\label{eq:solution_closed}
\psi_{m,q}(r,\phi) =J_m(k_{m,q}r) \begin{cases}
\sin(m\phi)\,, & -m\in\mathbb{N} \\
\cos(m\phi)\,, & +m\in\mathbb{N}_0 
\end{cases}\,,
\end{equation}
where $J_m$ are Bessel functions of the first kind.\footnote{We choose this labeling for convenience. 
The reader should not be confused by the fact that in the quantum mechanical interpretation the solutions \eqref{eq:solution_closed} are not eigenfunctions of the angular momentum operator, but only of the squared angular momentum operator whose eigenvalues do not depend on the sign of $m$.} 
The Dirichlet boundary conditions lead to the eigenvalues  $k_{m,q}=j_{m,q}/R$, where $j_{m,q}$ is the $q^{\mathrm{th}}$ zero of $J_m$,
$q \in \mathbb{N}$. Because the zeros of the Bessel functions do not depend on the sign of $m$
we have that the eigenvalues with $m \ne 0$ are doubly degenerate whereas those with
$m=0$ are not degenerate. Hence the eigenvalues of the Helmholtz equation with
Dirichlet boundary conditions are given by the multiset
\begin{equation} \label{eq:spectrum_closed_unptbd}
\sigma = \{ k_{m,q} = j_{m,q}/R: m \in \mathbb{Z}, \, q \in \mathbb{N}  \} \,.
\end{equation}
In the next section
we will also need  the Green function for the circular disk. It can be obtained
by taking the free Green function for the infinite plane and adding to it solutions of the Helmholtz
equation (\ref{eq:hholtz_qm}) such that the Dirichlet boundary conditions are satisfied.
The result is \cite{StewWaech1971}
\begin{eqnarray}\label{eq:green_func_closed}
G(\mathbf{r},\mathbf{r}',k) &=& -\frac{\ui}{4} \sum_{m=-\infty}^{\infty} \expp^{\ui m(\phi-\phi')}
J_m(kr_<) \left[H_m(kr_>) - \frac{H_m(kR)}{J_m(kR)} J_m(kr_>) \right] \nonumber \\
&=& -\frac{\ui}{4} H_0(k \vert \mathbf{r} - \mathbf{r}' \vert) + \frac{\ui}{4} \sum_{m=-\infty}^{\infty}
\expp^{\ui m(\phi-\phi')} \frac{H_m(kR)}{J_m(kR)} J_m(kr) J_m(kr') \,,
\end{eqnarray}
where here and in the following $H_m$ denotes the Hankel function $H_m^{(1)}$, and
$r_<$ and $r_>$ are the smaller and greater, respectively, of $r$ and $r'$.

As discussed in the previous section the trace formula (\ref{eq:trace_formula}) has the form
\begin{equation}\label{eq:trace_qm_closed}
d(k) = \sum_{m,q} \delta(k - k_{m,q}) \approx d_0(k) + d_{\mathrm{osc}}(k) \, ,
\end{equation}
where the sum is taken over the eigenvalues as described above. We will apply the
trace formula to investigate signatures of periodic orbits in the spectrum of the
cavity. For this purpose it is useful to consider the {\em length spectrum} which is
given by the Fourier transform $F(l)$ of the density of states. Since the trajectories
give oscillatory contributions of the form (\ref{eq:d_xi}) to the trace formula, $F(l)$
shows a peak structure at periodic orbits. In practice we can numerically determine
the spectrum only within a finite window $0<k<k_{\mathrm{max}}$. With this available
range in mind, the transform is weighted with a Gaussian cut-off function
$W(k)$ in order to reduce the influence of larger eigenvalues. We choose
$W(k)=\exp(-t k^2)$ with cut-off parameter $t=10/k_{\mathrm{max}}^2$. In our
numerical  evaluations $k_{\mathrm{max}}=100$ which corresponds to a window with 2456 eigenvalues.

Since $d(k)$ is a comb of delta functions its weighted Fourier transform is
\begin{equation} \label{eq:trace_qm_fourier_closed}
F(l) = \int_0^\infty d(k) \, W(k) \, \expp^{-\ui k l} \, \ud k = \sum_{m,q} \expp^{- \ui k_{m,q} l - k_{m,q}^2 t} \,,
\end{equation}
We will compare this result 
to its short-wave approximation which is obtained by evaluating the weighted Fourier transform
of the right-hand side of the trace formula (\ref{eq:trace_qm_closed}).

The smooth part of the density of states $d_0(k)$ is given by equation (\ref{eq:trace_smooth_closed})
with area $\mathcal{A}=\pi R^2$ and perimeter $L= 2 \pi R$. 
For brevity we will only consider the real part of the length spectrum. The real part of the weighted Fourier transform of the smooth part is then given by
\begin{equation}\label{eq:trace_smooth_fourier_closed}
\Real \, F_{0}(l) = \frac{\mathcal{A}}{4\pi t} - \frac{\expp^{-l^2 / 4t}}{8 \sqrt{\pi t}}
\left[L - \frac{\ui \mathcal{A} l}{t} \mathrm{erf} \left( \frac{\ui l}{2\sqrt{t}} \right) \right] \,,
\end{equation}
where $\mathrm{erf}(\cdot)$ is the error function.

\begin{figure}
\centerline{
\includegraphics[angle=0,width=5cm]{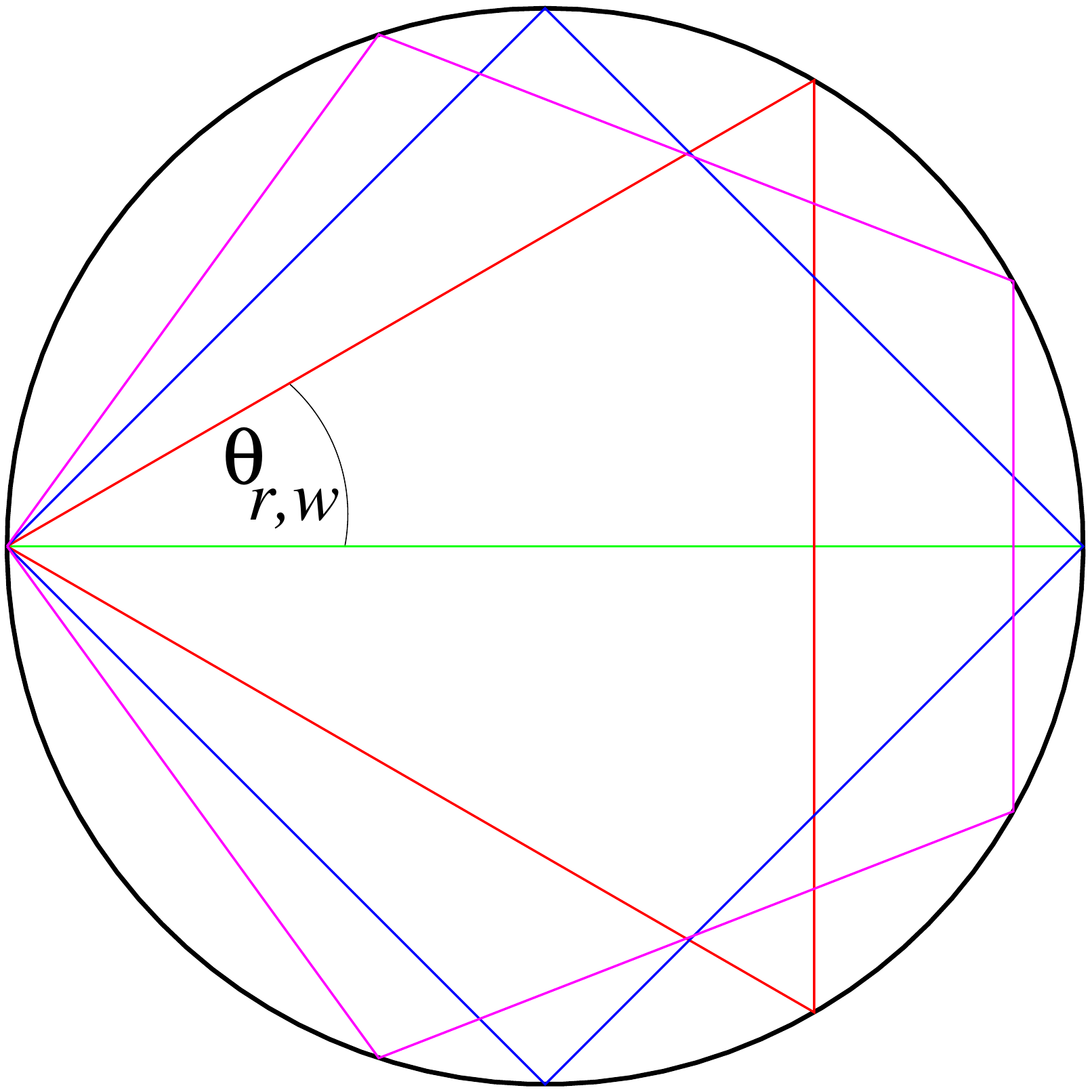} }
\caption[Periodic orbit families.] {\label{fig:periodic_orbs}
The four shortest periodic orbits in the circular cavity: the diameter, the equilateral triangle,
the square and the regular pentagon. All of the orbits occur in continuous families obtained
by rotation about the origin. The angle $\theta_{r,w}$ is shown for the triangular periodic orbit.}
\end{figure}

The oscillatory part of the density of states can be obtained by performing the 
trace in (\ref{eq:dos_green}) for the Green function of the circular disk
(\ref{eq:green_func_closed}) asymptotically as $k \rightarrow \infty$.
It is given by a sum over periodic orbits and has the form (see e.g.\ \cite{PalVatCser2001})
\begin{equation} \label{eq:trace_osc_closed}
d_{\mathrm{po}}(k) = \frac{2}{\pi} \sum_{r=2}^\infty \sum_{w=1}^{\lfloor r/2 \rfloor}
 g_p \, \mathcal{A}_p \, \sqrt{ \frac{2 k }{\pi l_p}} \,
\cos \left(k l_p - \nu_p \pi / 2 + \pi /4 \right) \, (-1)^r \,,
\end{equation}
where $\lfloor \cdot \rfloor$ is the floor function (integer part).
The periodic orbits in the circular disk occur in families because of the rotational symmetry
and we use the integers $r$ and $w$ to label the different families. For simplicity we 
combine them into a single index $p=(r,w)$. The orbits have $w$ rotations around the
billiard centre and $r$ reflections at the boundary. For $w=1$ these are regular polygons
with $r$ vertices. In Fig.~\ref{fig:periodic_orbs} we show representative
periodic orbits for some of the families with $w=1$. Each periodic orbit $\mathrm{PO}(r,w)$ 
reflects from the boundary with an angle of incidence $\theta_p$ measured from
the normal to the boundary, given by 
\begin{equation} \label{eq:ang_inci_po}
\theta_p = \frac{\pi}{2} - \frac{w \pi}{r} \,.
\end{equation}
We occasionally denote it by $\theta_{r,w}$. The periodic orbit length is 
\begin{equation} \label{eq:po_length}
l_p = 2 r R \cos(\theta_p) \,,
\end{equation}
and the area of the annulus filled by all the members of each family in configuration space is 
\begin{equation}\label{eq:po_area}
\mathcal{A}_p = \pi R^2 \cos^2(\theta_p) \,.
\end{equation}
The Maslov index $\nu_p$ in (\ref{eq:trace_osc_closed}) counts the number of conjugate
points and is simply given by $\nu_p=r$. The factor of $(-1)^r$ comes from the Dirichlet
reflections at the boundary as discussed earlier. Finally $g_p$ is a degeneracy factor
due to time-reversed orbits. We have $g_p=1$ for the diameter orbit and its repetitions and
$g_p=2$ for all other orbits.
We note that the sum over $r$ and $w$ in (\ref{eq:trace_osc_closed}) can be decomposed into a sum over relatively prime $r$ and $w$ and integer multiples of the two. 
The integer multiples then correspond to repetitions of a so called \emph{primitive periodic orbit}  characterized by the relatively prime $r$ and $w$.

\begin{figure}
\centerline{
\includegraphics[angle=0,width=10cm]{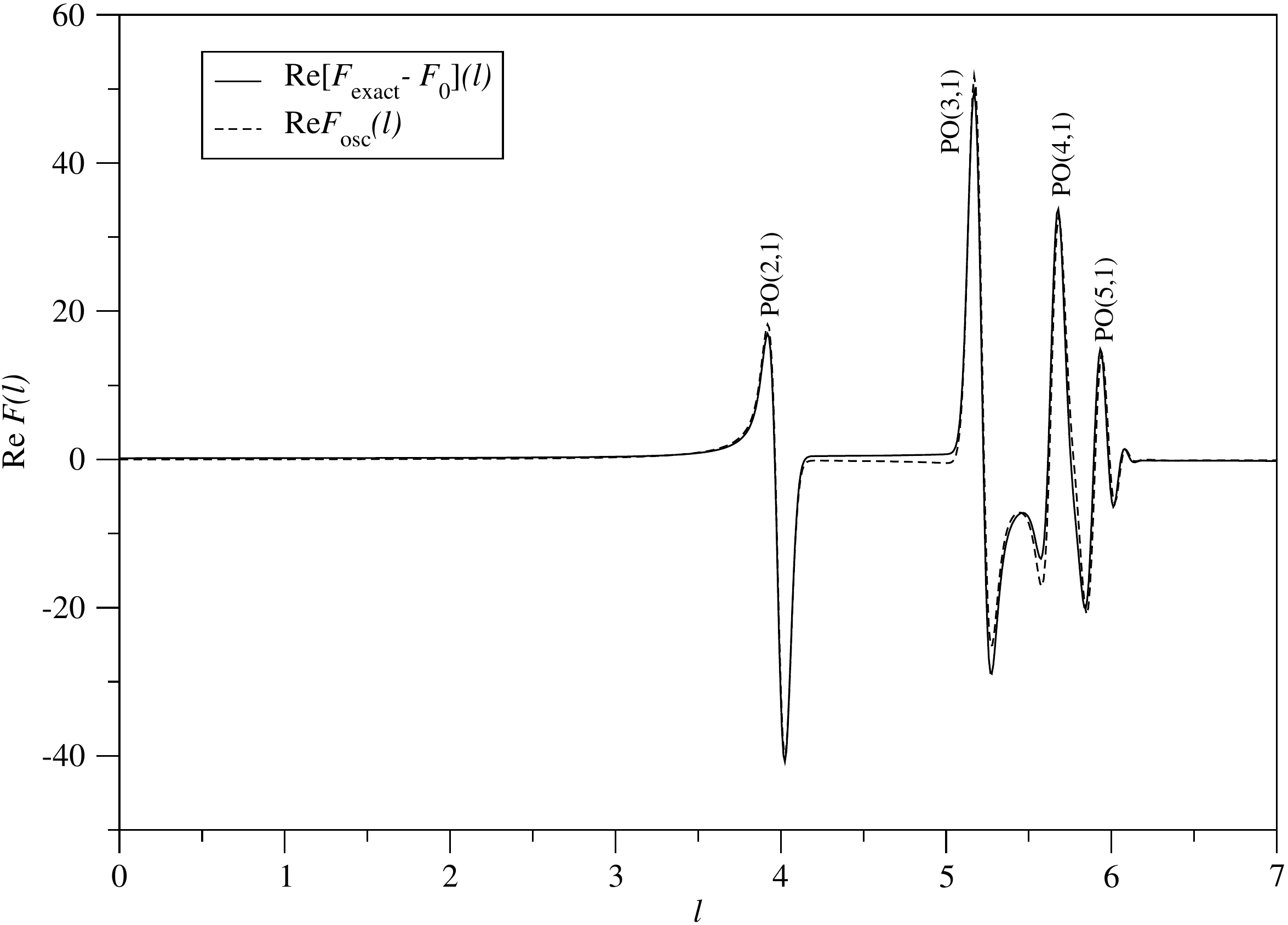}
}
\caption[Exact and semiclassical length spectra.]{\label{fig:length_spec_closed}
Length spectra for the unperturbed circular billiard. The exact spectrum is computed from
(\ref{eq:trace_qm_fourier_closed}) with unperturbed eigenvalues (minus the smooth part
(\ref{eq:trace_smooth_fourier_closed})) and the short-wave approximation from
(\ref{eq:trace_osc_fourier_closed}) with periodic orbits ($t=0.001$).
}
\end{figure}

The contribution of the periodic orbits to the length spectrum is obtained by taking the
Fourier transform with the weight function $W(k)$ of (\ref{eq:trace_osc_closed}). 
The integral can be performed analytically and its real part is given by
\begin{align} \label{eq:trace_osc_fourier_closed}
\Real F_{\mathrm{po}}(l) & = \frac{1}{2 \pi} \sum_p
\frac{g_p \, \mathcal{A}_p \, (-1)^{\lfloor (3 r+1)/2 \rfloor}}{ \sqrt{2 l_p}} \, \left( \frac{2}{t} \right)^{3/4}
\\ \notag & \times \left[ \expp^{-(l_p+l)^2/8 t} D_{1/2} \left( \frac{(l_p+l)}{\sqrt{2 t}} (-1)^{r+1}\right) +
\expp^{-(l_p-l)^2/8 t} D_{1/2} \left( \frac{(l_p-l)}{\sqrt{2 t}} (-1)^{r+1}\right) \right] \, ,
\end{align}
where $D_\nu$ denotes the parabolic cylinder function. In Fig.~\ref{fig:length_spec_closed},
we plot the oscillating part of the semiclassical length spectrum, $\Real F_{\mathrm{osc}}(l)$,
together with the exact length spectrum minus the smooth part $\Real[F(l) - F_{\mathrm{0}}(l)]$.
One sees that the short-wave approximation describes the exact curve very well, and it shows 
periodic orbits as peak structures at their corresponding lengths. Note that for fixed rotation
number $w$ we have $l_p \rightarrow 2\pi w$ as $r \rightarrow \infty$, and hence the periodic orbits approach
the limit of uniform circular motion around the inner edge of the billiard. The orbits therefore accumulate as
one approaches the values $l=2 w \pi$ from below in the length spectrum, and 
therefore the orbits $\mathrm{PO}(r,1)$ cannot all be resolved in Fig.~\ref{fig:length_spec_closed}.

\section{The closed cavity with point scatterer} \label{sec:ptbd_closed}
\subsection{Exact solutions} \label{sec:ptbd_closed_exact}

We perturb the system in the previous section by inserting a point scatterer. Without loss of generality we
locate it on the $x$-axis at a position $\mathbf{d}=(d,0)$ in Cartesian coordinates, see Fig.~\ref{fig:circle_cavity}.
Self-adjoint extension theory \cite{Zorbas1980} tells us that the perturbed eigenvalues of this system are solutions
of \cite{Shigehara1994}
\begin{equation} \label{eq:saet_cond}
1 - \lambda \, G_{\mathrm{reg}}(\mathbf{d},\mathbf{d},k) = 0 \,,
\end{equation}
where $\lambda$ is a coupling strength parameter and $G_{\mathrm{reg}}(\mathbf{d},\mathbf{d},k)$
is the regularised Green function.
The regularized Green function can be obtained from $G$ in (\ref{eq:green_func_closed}) 
by subtracting its divergent part. Replacing the Hankel  function $H_0$ in
(\ref{eq:green_func_closed})  by its asymptotic representation  as $\mathbf{r}' \rightarrow \mathbf{r}
= \mathbf{d}$ and subtracting the logarithmically divergent term gives
\begin{equation}\label{eq:green_reg_closed}
G_{\mathrm{reg}}(\mathbf{d},\mathbf{d},k) = -\frac{\ui}{4} + \frac{1}{2 \pi} \left( \log \frac{k}{2k_0}
+ \gamma \right) + \frac{\ui}{4} \sum_{m=-\infty}^{\infty} \frac{H_m(kR)}{J_m(kR)} J_m(kd) J_m(kd) \,,
\end{equation}
where  $\gamma$ is Euler's constant and $k_0$ is an arbitrary constant corresponding
to a choice of the regularisation. The imaginary part of (\ref{eq:green_reg_closed}) vanishes, and we
find that the perturbed eigenvalues, $\tilde{k}$, are determined as the solutions of $\tilde{f}(k)=0$,
where $\tilde{f}$ is the transcendental function
\begin{equation}\label{eq:quantcond_pted_closed}
\tilde{f}(k) = \frac{1}{2\pi} \left(\log {\frac{ka}{2} + \gamma}\right) - \frac{1}{4} \sum_{m=0}^{\infty}
\frac{Y_m(kR)}{J_m(kR)} \epsilon_m J^2_m(kd) 
\end{equation}
with
\begin{equation} \label{eq:epsilon_degen}
\epsilon_m = \begin{cases}
1\,, & m=0 \, , \\
2\,, & m \neq 0 
\end{cases}\,.
\end{equation}

The parameter $a$ in (\ref{eq:quantcond_pted_closed}) is a coupling constant which is a measure
of the strength of the perturbation. It replaces the previous parameters $\lambda$ and $k_0$ and
is related to them by $\lambda = -2 \pi / \ln (k_0 a)$. It has a direct physical interpretation in the
limit $a \rightarrow 0$ where the system with the point scatterer approaches that with a small scattering
disk of radius $a$ with Dirichlet boundary conditions \cite{RosWhelWir1996}. Further ways to
model the point scatterer experimentally will be discussed in section \ref{sec:ptbd_open_exact}.
The value of $a$ ranges from $0$ to $\infty$ which correspond to $\lambda\rightarrow0_+$ and
$\lambda\rightarrow0_-$, respectively. At these two limits the zeros of (\ref{eq:quantcond_pted_closed})
approach the unperturbed eigenvalues of Section~\ref{sec:cavity_closed}. 

The perturbation described by (\ref{eq:saet_cond}) is a rank one perturbation. In the case of degenerate
eigenvalues of the original problem this means that only one of the eigenvalues of a degenerate set
changes due to the perturbation while all others remain unaffected. For the unperturbed circle all
eigenvalues with ($m \neq 0$)  are two-fold degenerate, and only one of each pair changes according
to (\ref{eq:quantcond_pted_closed}) whereas the other remains as it is. This can also be seen
from the solutions (\ref{eq:solution_closed}), because the wavefunctions with angular dependence
$\sin(m\phi)$ have a nodal line on the $x$-axis and are unaffected by the presence of the scatterer. 
The spectrum $\tilde{\sigma}$ of all eigenvalues of the perturbed system is therefore
\begin{equation}\label{eq:spectrum_closed_ptbd}
\tilde{\sigma} = \{ \tilde{k}_{m,q} = k_{m,q} \in \sigma: -m, q \in\mathbb{N}\} \cup 
\{\tilde{k}_{m,q} \in \mathbb{R}: m \in \mathbb{N}_0, q \in \mathbb{N}, \tilde{f}(\tilde{k}_{m,q})=0\}\,,
\end{equation}
where we label the $\tilde{k}_{m,q}$ as follows.
We give them the same $m$ and $q$ numbers as the unperturbed
level at which they arrive as
$a \rightarrow \infty$. The reason for choosing the $a \rightarrow \infty$ and not the $a \rightarrow 0$
limit for the labelling is that otherwise we would miss the first perturbed level. This eigenvalue starts
at $k=k_{0,1}$ at $a=\infty$, and it decreases with decreasing $a$ and finally becomes complex.
A similar effect occurs also in one dimension where a delta function potential with negative coefficient
supports exactly one bound state \cite{Levin2002}.

\begin{figure}
\centerline{\includegraphics[angle=0,width=12cm]{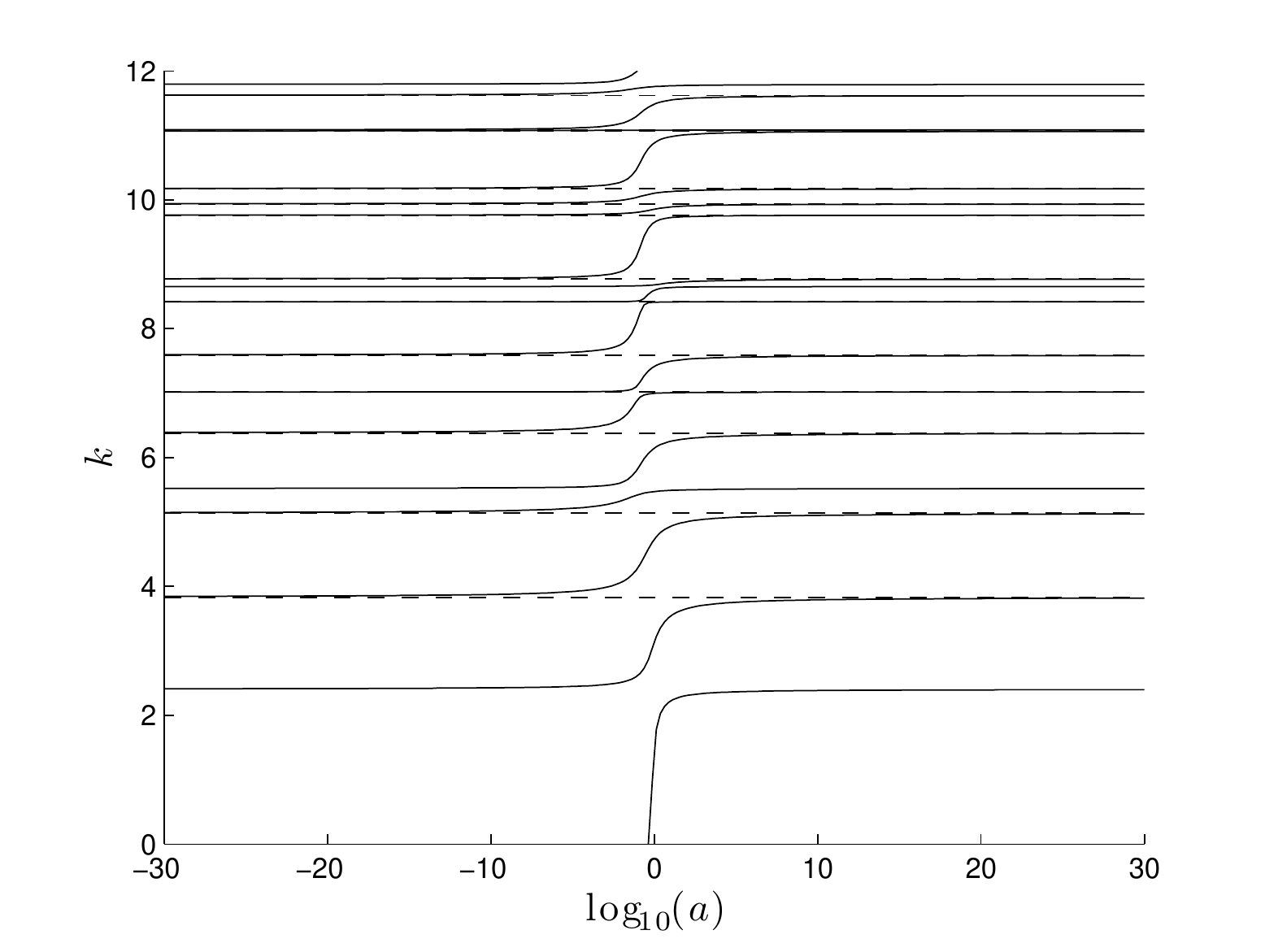}}
\caption[Level dynamics of the perturbed eigenvalues for $k=0...12$.]{\label{fig:leveldyn_closed}
Level dynamics of the perturbed eigenvalues of the circular cavity with point scatterer positioned at $d=0.59$ for $k \in [0,12]$.
The solid curves are the eigenvalues which change and the dashed horizontal lines are the eigenvalues which do not
change as $a$ is varied.}
\end{figure}

As an illustration we plot in Fig.~\ref{fig:leveldyn_closed} the eigenvalues of a circular disk with point scatterer at $d=0.59$
as a function of the parameter $a$. They have the property that there is always one perturbed eigenvalue between any two
adjacent unperturbed ones. This follows from the fact that the Green function is a monotonically decreasing function between
its poles. We used this property to solve (\ref{eq:saet_cond}) using a bisection method. Due to the logarithmic dependence on
the strength in (\ref{eq:quantcond_pted_closed}), we vary $a$ exponentially from $a=10^{30}$ to $a=10^{-30}$. Note that
the first perturbed eigenvalue is complex beyond a certain value of $a$.
 
\begin{figure}
\centerline{
\includegraphics[angle=0,width=10cm]{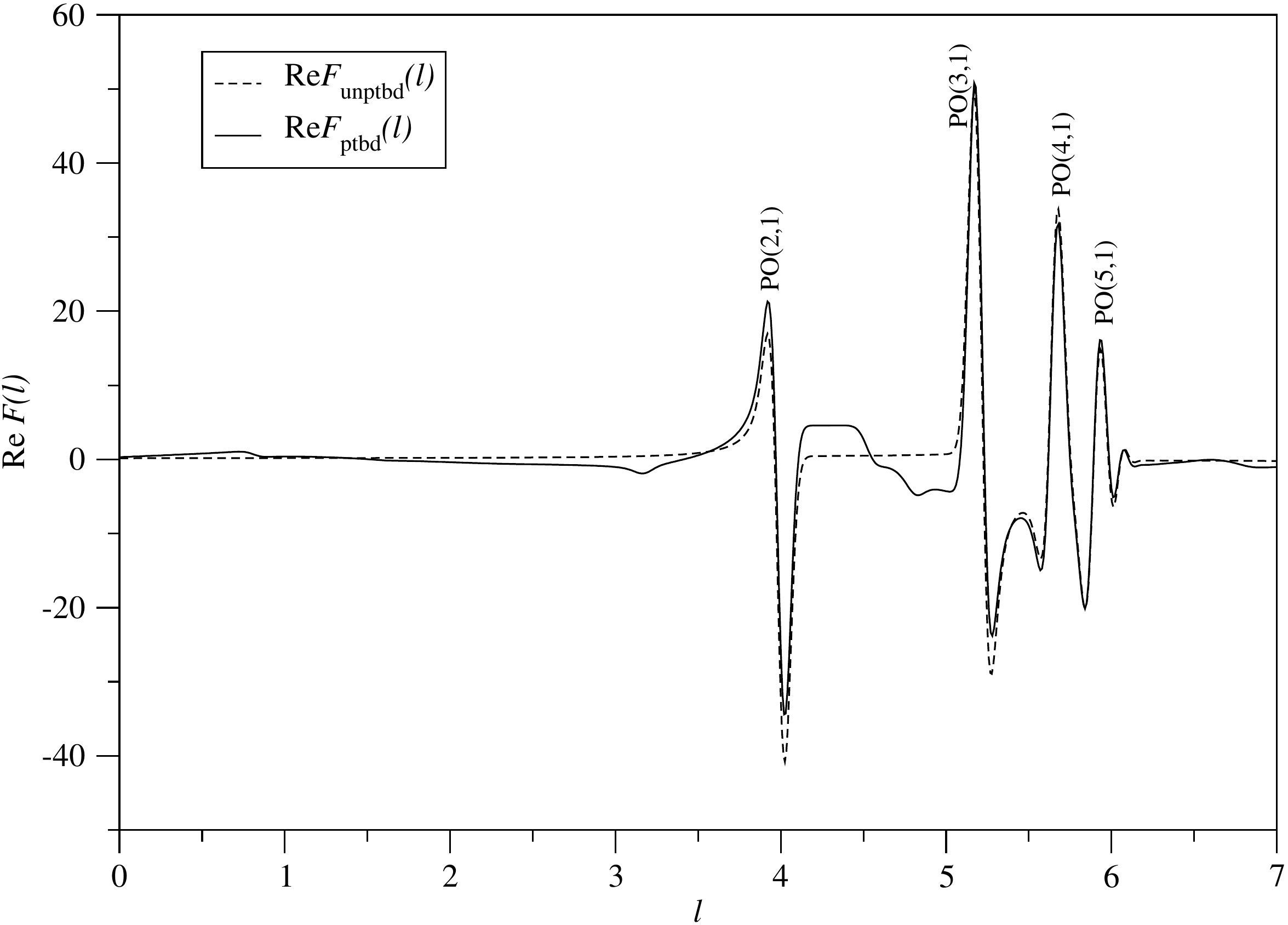}
}
\caption[Unperturbed and perturbed exact length spectra.]{\label{fig:length_spec_unptbd_ptbd_closed}
Exact length spectra for the unperturbed and perturbed circular cavities. The unperturbed spectrum is computed
from (\ref{eq:trace_qm_fourier_closed}) and the perturbed spectrum from (\ref{eq:trace_qm_fourier_closed_ptbed}).
Both spectra have had the smooth part subtracted ($d=0.59$, $a_0=1.0$, $t=0.001$).
}
\end{figure}

We consider now the length spectrum (\ref{eq:trace_qm_fourier_closed}) of the perturbed cavity. It is of the same form as before, namely 
\begin{equation} \label{eq:trace_qm_fourier_closed_ptbed}
\tilde{F}(l) = \sum_{m,q} \expp^{-\ui l \tilde{k}_{m,q} - t \tilde{k}_{m,q}^2} \,,
\end{equation}
where the sum is now over the eigenvalues of the perturbed disk. The parameter $t$ is again $t=10/k_{\mathrm{max}}^2$ with
$k_{\mathrm{max}} = 100$. In Fig.~\ref{fig:length_spec_unptbd_ptbd_closed} we plot the unperturbed and perturbed length
spectra together, with the smooth part $F_0(l)$ in (\ref{eq:trace_smooth_fourier_closed}) subtracted. We see that the
periodic orbit structures are the same for both curves, but there also are  small discrepancies. We will see in the next
section that these discrepancies can be attributed  to diffractive orbits. In order to visualize their contribution better we will consider in
the following the difference, $\Delta F(l)=\tilde{F}(l) - F(l)$, of the perturbed and unperturbed length spectra. 

%%%%%%%%%%%%%%%%%%%%%%%%%%%%%%%%%
\subsection{Short-wave approximation} \label{sec:ptbd_closed_sc}

The short wavelength approximation for the perturbed system can be obtained again from its Green function.
This Green function follows from self-adjoint extension theory as
\begin{equation} \label{eq:green_ptbd}
\widetilde{G}(\mathbf{r},\mathbf{r}',k) = G(\mathbf{r},\mathbf{r'},k) + G(\mathbf{r},\mathbf{d},k)
\frac{\lambda}{1 - \lambda G_{\mathrm{reg}}(\mathbf{d},\mathbf{d},k)} G(\mathbf{d},\mathbf{r}',k) \, ,
\end{equation}
where $G$ is the Green function of the unperturbed disk. If one then evaluates the trace of the Green function
in (\ref{eq:dos_green}) to obtain the density of states asymptotically as $k \rightarrow \infty$, one finds that
the first term on the right hand side of (\ref{eq:green_ptbd}) gives the same contributions as in the unperturbed
case (mean part and periodic orbit terms). The second term gives additional contributions. If one expands the
fraction in a geometric series one finds that these contributions come from trajectories that start and end at
the scatterer and do this an arbitrary number of times. These are the diffractive trajectories that occur
in Keller's geometrical theory of diffraction \cite{Keller1962}. 

For a single diffractive orbit, labelled by $d$, the contribution to the density of states is \cite{VatWirRos1994}
\begin{equation}\label{eq:trace_diffrac}
d_{d}(k) = \Real \left\{ \frac{l_d}{\pi} \left[\prod_{j=1}^{\mu_{d}} \frac{\mathcal{D}(k)}{\sqrt{8 \pi k \vert(M_{d,j})_{12} \vert}}
\right] \exp \left[\ui \left(k L_d - \frac{\nu_d \pi}{2} - \frac{3 \mu_d \pi}{4} \right) \right] (-1)^{r_d} \right\} \,,
\end{equation}
where $j$ labels the different parts of the diffractive orbit between consecutive encounters with the scatterer.
The total length of the diffractive orbit is $L_d$, the Maslov index $\nu_d$ counts the number
of conjugate points, $r_d$ is the total number of reflections with the boundary and 
$(M_{j})_{12}$ is the 12-element of the monodromy matrix $M_j$. The total number of encounters of
a diffractive orbit with the scatterer is $\mu_d$. Furthermore, the length $l_d$ is the total length $L_d$,
divided by the repetition number in the case that the diffractive orbit is a multiple repetition of a shorter one.
The diffraction coefficient is $\mathcal{D}$, which for a point scatterer is independent of
incoming and outgoing angles and  given by \cite{RosWhelWir1996,ExnerSeba1996}
\begin{equation}\label{eq:diff_const}
\mathcal{D}(k)  = \frac{2 \pi}{\ui \pi / 2 - \gamma - \log(ka/2)} \,.
\end{equation}
Finally we should mention that if a trajectory goes straight through the scatterer without changing its angle,
then it counts twice, once counting the point of coincidence with the scatterer as encounter and once not
counting it as such. In our computations we chose a value of $a=1$ for our perturbation parameter.
This value is suitable for seeing clear diffractive effects for wavenumbers in the range $[0,100]$.

\begin{figure}
\centerline{
\includegraphics[angle=0,width=8cm]{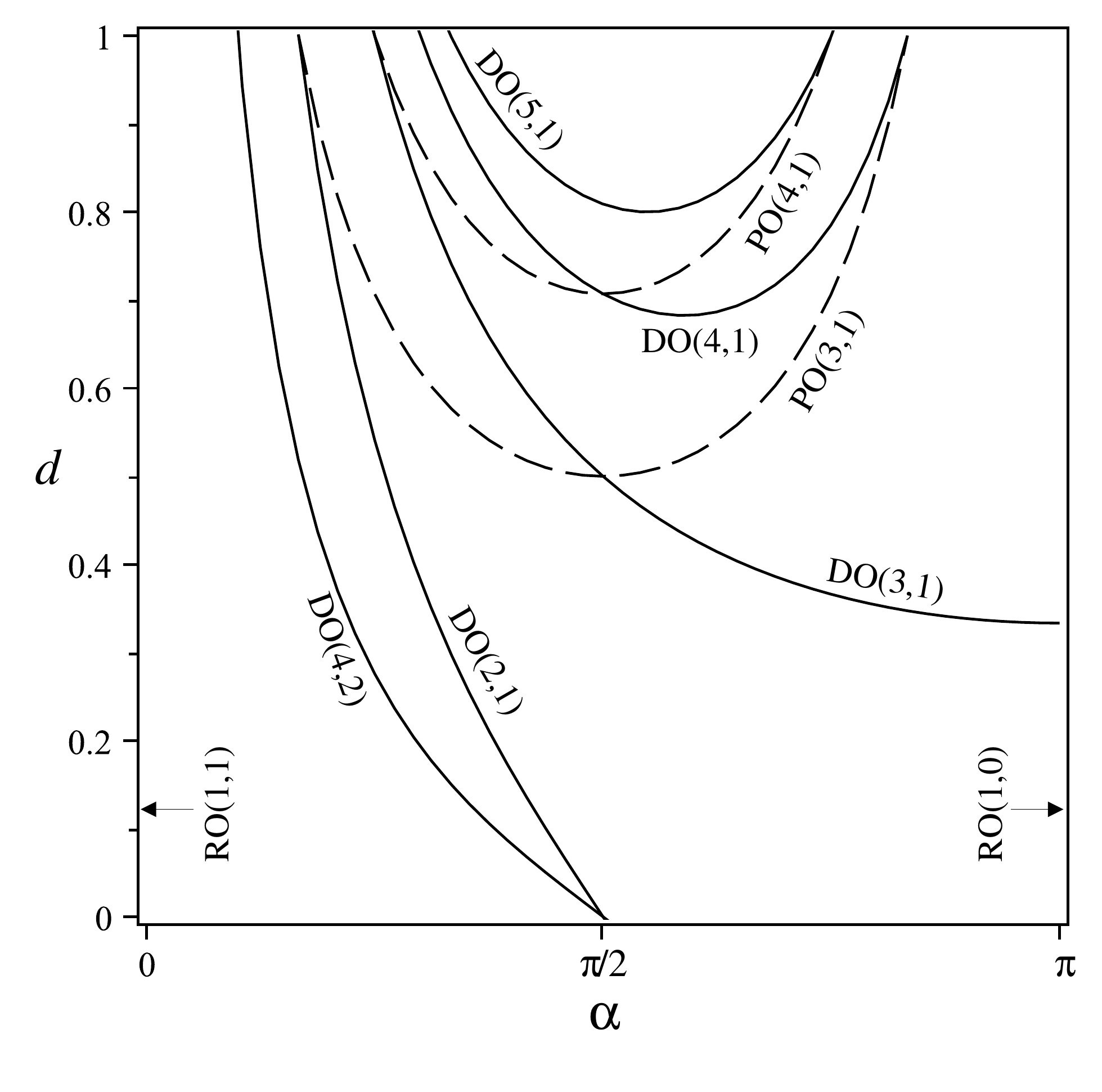}
}
\caption[Point scatterer position as a function of the diffractive orbit initial angle.]{\label{fig:diffrac_angles}
Point scatterer position $d$ as a function of the initial angles $\alpha$ of diffractive orbits $\mathrm{DO}(r,w)$.
For fixed $r$ and $w$, the function (\ref{eq:difforb_cond_alpha}) is plotted  (solid lines) for the diffractive orbits
and the function (\ref{eq:difforb_cond_alpha_po}) is plotted (dashed lines) for the periodic orbits. The radial
diffractive orbits $\mathrm{RO}(2k+1,k+1)$ and $\mathrm{RO}(2k+1,k)$ correspond to the vertical lines
$\alpha=0$ and $\alpha=\pi$, respectively.}
\end{figure}

We investigate now the different kinds of diffractive orbits that occur in the circular cavity. It is sufficient
to consider diffractive orbits that have just one encounter with the scatterer, because all other diffractive
orbits can be obtained from combinations of them. These single encounter diffractive orbits occur in two
forms: they do or do not coincide with a periodic orbit that goes through the scatterer. We investigate first the
latter type and generically label them as $\mathrm{DO}(r,w)$, where $r$ is again the number of reflections
at the boundary and $w$ is the number of rotations around the billiard centre. Each diffractive orbit has $r+1$
legs. This gives the angle of incidence $\theta$ of diffractive orbits as solutions of the transcendental equation \cite{Brack2009}
\begin{equation}\label{eq:difforb_cond_theta} 
d \cos \left( \frac{r \pi}{2} - r \theta \right) = (-1)^w R \sin \theta \,.
\end{equation}
The simple dependence on the rotation number $w$ in the condition (\ref{eq:difforb_cond_theta})
means that the diffractive orbits are separated into two classes; one class consists of orbits with an odd number of
rotations, the other of orbits with an even number. It is more convenient to write the condition in terms of the initial
angle $\alpha$ at which each diffractive orbit starts with respect to the symmetry axis. The relation between
this angle and the angle of incidence is
\begin{equation} \label{eq:angle_relation}
\theta = \frac{1}{r} \left( \frac{(r-1) \pi}{2} - (w-1) \pi - \alpha \right)\,.
\end{equation}
This gives the position of the scatterer as a function of the initial angle as
\begin{equation}\label{eq:difforb_cond_alpha} 
d = \frac{R}{\sin\alpha} \sin \left( \frac{(r-1) \pi}{2 r} - \frac{\alpha}{r} - \frac{(w-1) \pi}{r} \right)\,.
\end{equation}
In Fig.~\ref{fig:diffrac_angles} we plot the condition (\ref{eq:difforb_cond_alpha}) for a few orbits that we will later use.
For a particular choice of $r$, $w$ and $d$, this condition is solved numerically for $\alpha$ by a Newton procedure.
Depending on the choice of $d$, we see that in some instances there is no solution for the diffractive orbit and in
others there are two solutions. In this way, some orbits are created by tangent bifurcations as the parameter $d$
is varied and we will discuss this later.

\begin{figure}
\centerline{
\includegraphics[angle=0,width=8cm]{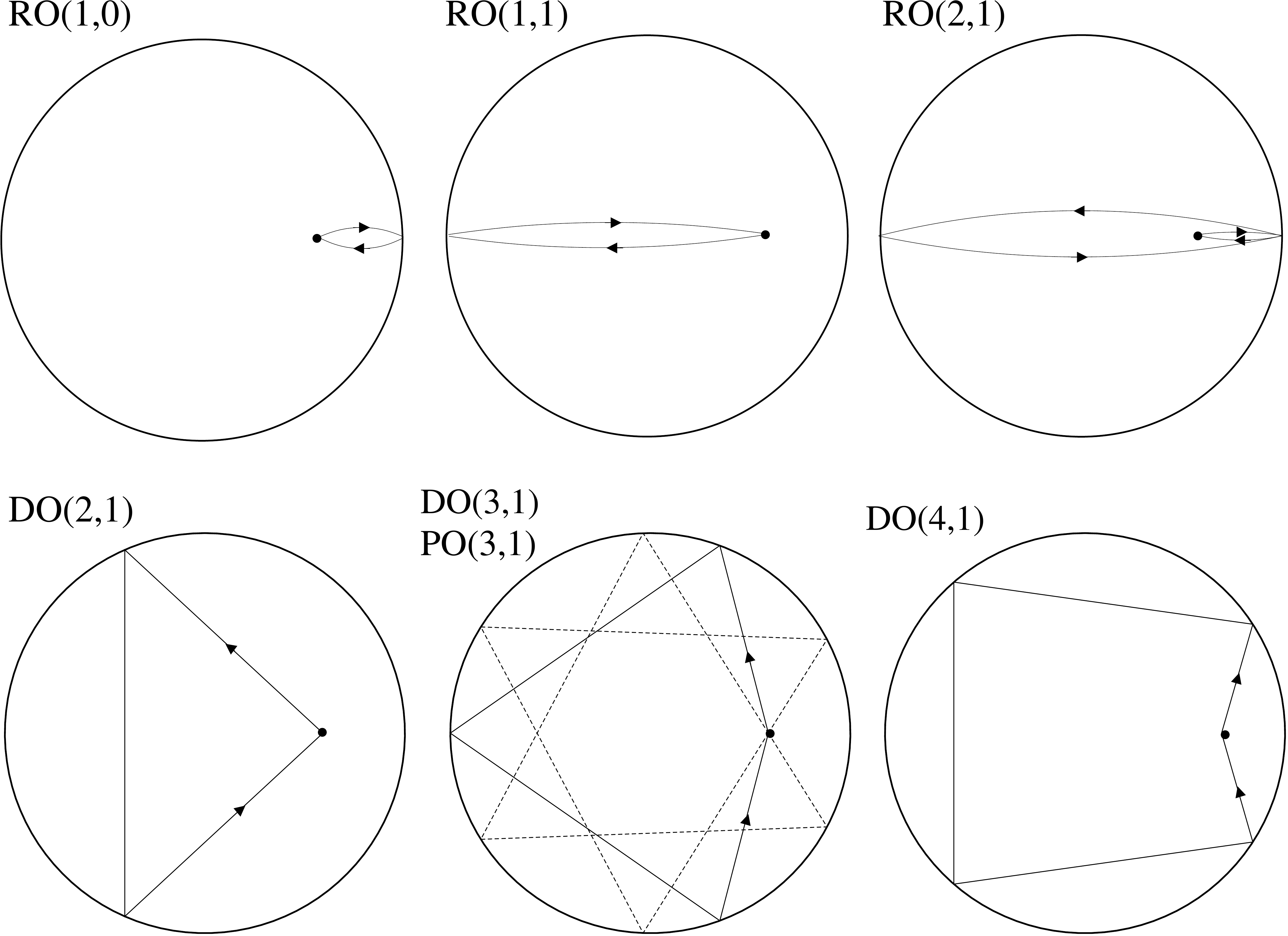}
}
\caption[Diffractive orbits.]{\label{fig:diffractive_orbits}
The first few diffractive orbits by length of the circular billiard with point scatterer. The scatter is positioned at $d=0.59$
in all panels except for $\mathrm{DO}(4,1)$, where it is positioned at $d\approx$0.6825, the bifurcation point where the
real pair of orbits $\mathrm{DO}(4,1)$ and $\mathrm{DO}(4,1)'$ come into existence. For $\mathrm{DO}(3,1)$, we also
show the corresponding diffractive periodic orbit $\mathrm{PO}(3,1)$, with its reflected partner (dashed lines),
which comes into existence at $d=0.5$.
}
\end{figure}

There are two radial diffractive orbits in the circle billiard both travelling along the $x$-axis and both having $r=1$ reflections
with the boundary. The longer of these two orbits passes through the centre of the billiard and we hence label it as
$\mathrm{RO}(r,w)=\mathrm{RO}(1,1)$. The shorter orbit does not pass through the centre and we hence label it as
$\mathrm{RO}(r,w)=\mathrm{RO}(1,0)$. Either of these two orbits can be added to multiples of the radial periodic orbit
$\mathrm{PO}(2,1)$ where we regard  $\mathrm{PO}(2,1)$ as always missing the scatterer. In this way, these combination
orbits are primitive as they always only contain one encounter with the scatterer. We therefore denote all these orbits by
$\mathrm{RO}(2k+1,k+1)$ and $\mathrm{RO}(2k+1,k)$ for $k \in \mathbb{N}_0$, for the longer and shorter orbits respectively.

The first and last legs of each diffractive orbit are of equal length 
\begin{equation}
x = d \cos \alpha + R \cos \theta\,. 
\end{equation}
The remaining $r-1$
legs are of length $l = 2 R \cos \theta$, giving the total length of the orbit as 
\begin{equation}\label{eq:diff_orb_length}
l_d = 2 (r R \cos \theta + d \cos \alpha).
\end{equation}
Using the basic monodromy matrices for motion on a straight path and reflection at a curved boundary \cite{Sieber1990}
we find the 12-element of the monodromy matrix to be
\begin{equation} \label{eq:monodromy_12}
M_{12} = \frac{2}{R \cos\theta}(R \cos\theta - x)((r-1) R \cos \theta - r x)\,.
\end{equation}
The Maslov index $\nu_d$ in (\ref{eq:trace_diffrac}) is the number of conjugate points and can be computed from the number
of times $M_{12}$ changes sign along an orbit path. The result is
\begin{equation}
\nu_d = \left\{ \begin{array}{cr}
r-2\,, & x < (l - l/r)/2 \\
r-1\,, & (l - l/r)/2 < x < l/2 \label{eq:conj_points_do}\\
r\,, & x > l/2
\end{array} \right.\,,
\end{equation}
which also remains valid for the radial orbits $\mathrm{RO}(1,0)$ and $\mathrm{RO}(1,1)$ even though the length $l=2R$
does not form part of these two orbits. Due to the time-reversal symmetry in the billiard system each diffractive orbit comes
in a pair, except for the radial orbits RO of which there is only one of each.

As mentioned before, diffractive orbits may also occur in the form of periodic orbits, and we denoted them also by
$\mathrm{PO}(r,w)$, depending on whether the condition
\begin{equation}\label{eq:caus_radius}
R \sin \theta = R \cos \left( \frac{w\pi}{r} \right) \leq d
\end{equation}
holds, where $r$ is again the number of encounters with the boundary and $w$ the number of rotations around the disk's centre.
For a given periodic orbit family, for which each member has an angle of incidence at the boundary given by (\ref{eq:ang_inci_po}),
the value of $d$ at which we have equality in (\ref{eq:caus_radius}) is the caustic radius of the family. There are two isolated
members of each family that serve as diffractive orbits which are related to each other by symmetry with respect to the $x$-axis.
For both these orbits, the first and last legs have lengths $x_{1,2}=(l \pm \sqrt{l^2 + 4(d^2 - R^2)})/2$, respectively, where we
again have $l = 2 R \cos \theta$ and the 12-element of the monodromy matrix for diffractive-periodic orbits is given by
\begin{equation}\label{eq:monodromy_periodic}
M_{12} = \frac{2 r}{R \cos\theta} \left( R^2 \sin^2 \theta - d^2 \right)\,.
\end{equation}
The corresponding number of conjugate points is $\nu_d = r - 1$. Each of the periodic orbits is doubly degenerate due to the reflection
symmetry and in turn each of these is doubly generate due to time reversal symmetry. Each primitive periodic orbit must therefore
be counted four times.

Similarly to the diffractive orbits, we can give an expression for the position of the scatterer in terms of the initial angle $\alpha$
at which each periodic orbit starts. It is given by
\begin{equation}\label{eq:difforb_cond_alpha_po}
d = R \frac{\sin \theta}{\sin \alpha} \,.
\end{equation}
We plot this condition in Fig.~\ref{fig:diffrac_angles} alongside the corresponding curves for the diffractive orbits for a couple
of orbits that we will use later. We see that the curves corresponding to the periodic orbits intersect certain curves corresponding to
the diffractive orbits and in this way some orbits are created by pitchfork bifurcations as the parameter $d$ is varied. We will give
a review of all the bifurcations later.

We have so far only considered diffractive orbits which have one encounter with the scatterer. In this case we have $\mu_d=1$
and $l_d=L_d$ in (\ref{eq:trace_diffrac}). Diffractive orbits consisting of more than one encounter with the scatterer are simply
combinations of the single encounter orbits that we have discussed. It is important to distinguish between multiply diffractive
radial orbits that are combinations of the primitive orbits $\mathrm{RO}(1,0)$, $\mathrm{RO}(1,1)$ and $\mathrm{PO}(2,1)$
and the singly-diffractive orbits $\mathrm{RO}(2k+1,k)$ and $\mathrm{RO}(2k+1,1+k)$. The diffractive contribution to the
trace formula for the circular billiard with point scatterer is finally

\begin{equation}\label{eq:diff_conbution}
d_{\mathrm{do}}(k) = \sum_{d} A_d (-1)^{r_d} \exp \left\{ \ui \left(k L_d - \frac{\nu_d \pi}{2}
- \frac{3 \mu_d \pi}{4} \right) \right\} + \mathrm{c.c.} \,,
\end{equation}
where the amplitudes are given by
\begin{equation}\label{eq:amplitude}
A_d = \frac{l_d}{2 \pi} \left \{ \prod_{j=1}^{\mu_d} \frac{g_{d,j} \mathcal{D}(k)}{\sqrt{8 \pi k \vert(M_{d,j})_{12} \vert}} \right\} \, ,
\end{equation}
and the sum runs over all diffractive orbits described above, labelled by $d$. The product runs over all primitive diffractive orbits
$\mathrm{RO}(r,w)$, $\mathrm{DO}(r,w)$ and $\mathrm{PO}(r,w)$. In the amplitudes (\ref{eq:amplitude}) we have also included
a degeneracy factor $g_{d,j}$ which takes into account the double degeneracy of non-radial diffractive orbits and the quadruple degeneracy
of the non-radial periodic orbits, as discussed above, i.e.
\begin{equation}
g_{d,j} = \left \{ \begin{array}{ll}
1\,, & \textrm{if $(d,j) \corresponds$ RO$(2k+1,k)$, RO$(2k+1,1+k)$, $k\in\mathbb{N}_0$} \\
2\,, & \textrm{if $(d,j) \corresponds$ DO$(r,w)$, PO$(2,1)$} \\
4\,, & \textrm{if $(d,j) \corresponds$ PO$(r,w)$, $r\ge3$}
\end{array} \right. \,.
\end{equation}

The presence of the scatterer also adds a small additional term to the smooth part $d_0(k)$ of the density of states. It can be
obtained from the Green function (\ref{eq:green_ptbd}). The smooth part is then given by
\begin{equation}\label{eq:trace_smooth_closed_ptbd}
d_{0}(k) = \frac{\mathcal{A} k}{2 \pi} - \frac{L}{4 \pi} - \frac{1}{2 k}
\left[ \frac{\pi^2}{4} + \left(\gamma + \log \frac{ka}{2} \right)^2 \right]^{-1} \,.
\end{equation}

For convenience, we slightly change the definition of the length spectrum for the perturbed cavity and consider
\begin{equation} \label{eq:trace_sc_fourier_closed}
F(l) = \Real \int_0^\infty k^{\mu_{\mathrm{max}/2}} d(k) \, W(k) \expp^{- \ui k l} \ud k \,,
\end{equation}
where we introduce $\mu_{\mathrm{max}}$ as the chosen maximum number of encounters of all the
diffractive orbits with the scatterer to be included in the short wavelength approximation. This removes the singularity
in the integrand when $d(k)$ is replaced by its approximation. The weight function $W(k)$ is the same as
in Section \ref{sec:cavity_closed}. In order to see the effect of the diffractive orbits in the perturbed spectrum
we then consider the difference between the perturbed and unperturbed length spectra $\Delta F(l) = \tilde{F}(l) - F(l)$.
In the short wavelength approximation this quantity is given only by the diffractive orbit contribution and the extra
contribution to the smooth part, because the periodic orbit terms and the area and perimeter terms in the smooth
part cancel. We hence have
\begin{equation}\label{eq:trace_sc_fourier_closed_diff}
\Delta F(l) \approx \int_0^\infty k^{\mu_{\mathrm{max}/2}} \left \{d_{\mathrm{do}}(k) - \frac{1}{2 k}
\left[ \frac{\pi^2}{4} + \left(\gamma + \log \frac{ka}{2} \right)^2 \right]^{-1} \right\} W(k) \expp^{-\ui k l} \ud k \,.
\end{equation}
This integral will be computed numerically. In the lower panel of Fig.~\ref{fig:length_spec_unptbd_ptbd_closed} we plot
$\Delta F(l)$ for a scatterer position $d=0.59$, scatterer strength $a=1$ and $\mu_{\mathrm{max}}=4$ (full line)
and compare it to the approximation (\ref{eq:trace_sc_fourier_closed_diff}) (dashed line). The diffractive orbits included
are those given in Fig.~\ref{fig:diffractive_orbits}, except $\mathrm{DO}(4,1)$ and $\mathrm{DO}(4,1)'$ which are not real
at $d=0.59$ (we detail this below), plus the radial periodic orbit $\mathrm{PO}$(2,1). We see that the peaks can be
identified with the contributions of the diffractive orbits. The agreement is very good except at a few particular lengths.

\begin{figure}
\centerline{
\includegraphics[angle=0,width=12cm]{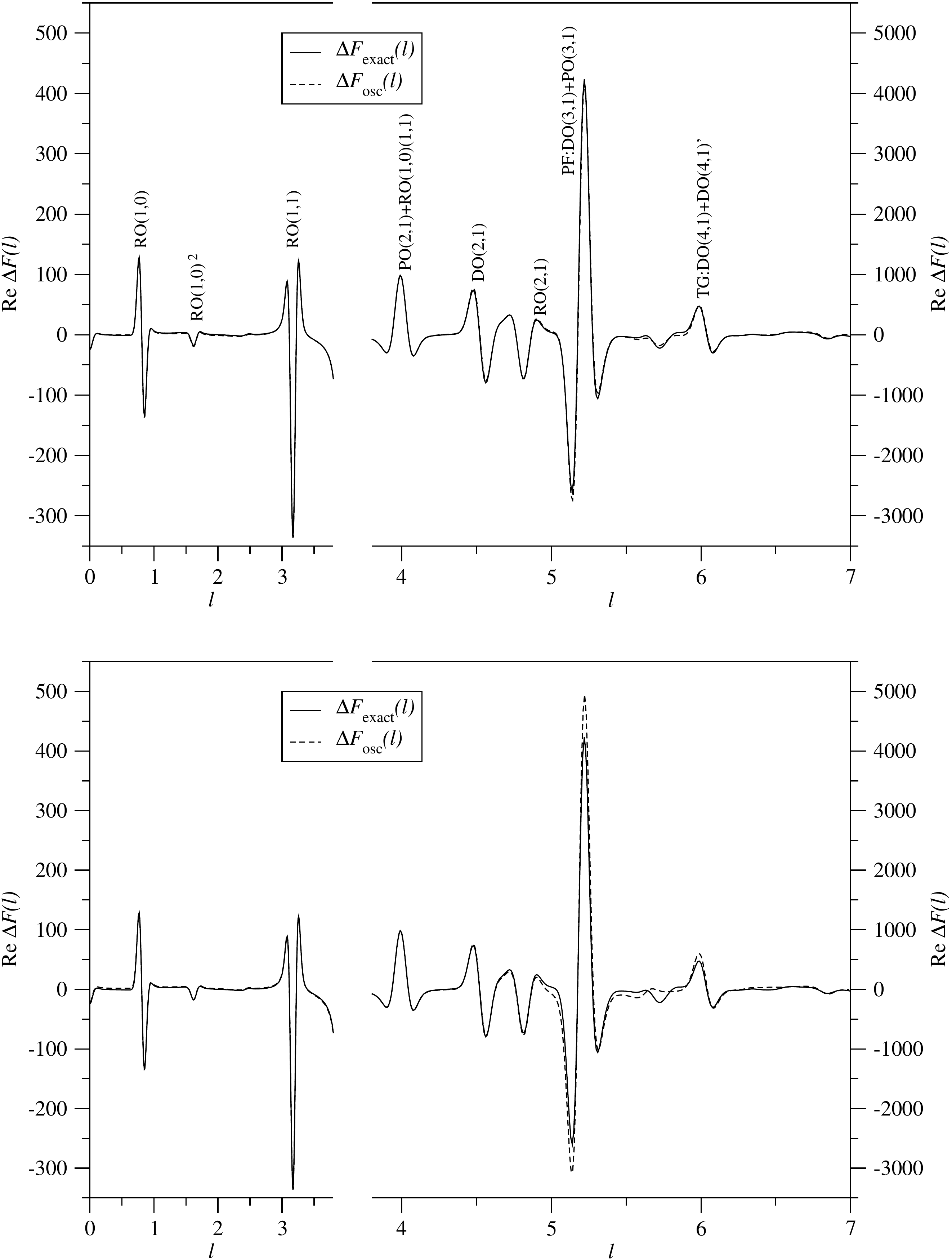}
}
\caption[Exact and semiclassical length spectrum differences.]{\label{fig:length_spec_diff_closed}
Length spectra difference $\Delta F(l)$ for the perturbed circular billiard. The exact difference (full lines) is computed from
unperturbed and perturbed eigenvalues, and the short-wave approximation (dashed lines) from
(\ref{eq:trace_sc_fourier_closed_diff}) with diffractive orbits. The upper panel includes uniform approximations
for the $\mathrm{DO}(3,1)+\mathrm{PO}(3,1)$ pitchfork bifurcation (PF) and the $\mathrm{DO}(4,1) +  \mathrm{DO}(4,1)'$
tangent bifurcation (TG). These are labelled on the plot next to the corresponding peak structure as are the other contributing
orbits. The lower panel shows the short-wave approximation without uniform approximations. ($d=0.59$, $a=1.0$,
$\mu_{\mathrm{max}}=4$, $t=0.001$.)
}
\end{figure}

We have already mentioned that the diffractive orbits and isolated periodic orbits that serve as diffractive orbits are involved
in pitchfork and tangent bifurcations. As with standard periodic orbit contributions, the diffractive contribution (\ref{eq:diff_conbution})
to the trace breaks down near bifurcations of the diffractive orbits, because the matrix element $M_{12}$ vanishes there.
If one chooses a value of $d$ in such a way that the scatterer is away from bifurcations then the length spectra
difference $\Delta F$ is well described by the approximation (\ref{eq:trace_sc_fourier_closed_diff}). In order to deal with
all values of $d$, one has to apply uniform approximations to the diffractive contribution to the density of states. 
The bifurcation scenarios for closed orbits (diffractive orbits) in a circular billiard have recently been described in \cite{Brack2009}
and we here summarise the results in Tab.~\ref{tab:bifurcations}.

The bifurcations occur in three groups. Group I consists of diffractive orbits $\mathrm{DO}(2k,k)$ (even $r$ and maximum $w$)
that are created from the break-up of the tori (BT) at $d=0$ of the radial diffractive periodic orbits $\mathrm{PO}(2k,k)$.
Group II consists of diffractive orbits $\mathrm{DO}(2k+1,k)$ (odd $r$ and maximum $w$) that are created by pitchfork (PF) bifurcation
at $d=R/r$ from the radial diffractive orbits $\mathrm{RO}(2k+1,k)$. These orbits undergo a further PF bifurcation at
$d=R\cos(\theta_{r,w})$ to create the periodic orbits $\mathrm{PO}(2k+1,k)$ (note that this value of $d$ is the caustic
radius of the torus which contains the created periodic orbit). Group III consists of all other diffractive orbits. They are created in pairs
$\mathrm{DO}(r,w)$ and $\mathrm{DO}(r,w)'$ by tangent (TG) bifurcation at $d_{r,w}$, which corresponds to the value
of each minimum in Fig.~\ref{fig:diffrac_angles}. One member of each of these pairs, $\mathrm{DO}(r,w)$, undergoes
a PF bifurcation, again at the caustic radius $d=R\cos(\theta_{r,w})$, to create the periodic orbit $\mathrm{PO}(r,w)$.
For group III, both orbits are not real before the bifurcation (leading to so called \emph{ghost orbits}, see below). At the boundary, all the DOs become POs as indicated
in the last column. Finally, we note that for the circle billiard there are no bifurcations involving only periodic orbits. 

\begin{table*}[htbp]
\begin{center}
\small
\begin{tabular}{cc|c|c|cc|}
\cline{3-6}
\phantom{ } & \phantom{ } & \multicolumn{4}{|c|}{bifurcation sequence group}\\\cline{3-6}
\phantom{ } & \phantom{ } & 
I: $\begin{array}{c}r\textrm{ even}\\w=r/2\end{array}$ &
II: $\begin{array}{c}r\textrm{ odd}\\w=(r-1)/2\end{array}$ &
\multicolumn{2}{|c|}{III: $\begin{array}{c}r\textrm{ even/odd}\\w<(r-1)/2\end{array}$}\\\hline
\multicolumn{1}{|c|}{\multirow{4}{*}{1st bifurcation}} & type & BT & PF & \multicolumn{2}{|c|}{TG} \\\cline{2-6}
\multicolumn{1}{|c|}{\phantom{ }} & at $d=$ & $0$ & $R/r$ & \multicolumn{2}{|c|}{$d_{r,w}$} \\\cline{2-6}
\multicolumn{1}{|c|}{\phantom{ }} & from & $\mathrm{PO}(r,w)$ & $\mathrm{RO}(2k+1,k)$ & \multicolumn{2}{|c|}{$-$} \\\cline{2-6}
\multicolumn{1}{|c|}{\phantom{ }} & to & $\mathrm{DO}(r,w)$ & $\mathrm{DO}(r,w)$ &
\multicolumn{1}{|c|}{$\mathrm{DO}(r,w)'$}  & $\mathrm{DO}(r,w)$ \\\hline
\multicolumn{1}{|c|}{\multirow{3}{*}{2nd bifurcation}} & type & $-$ & PF & \multicolumn{1}{|c|}{$-$} & PF \\\cline{2-6}
\multicolumn{1}{|c|}{\phantom{ }} & at $d=$ & $-$ & $R\cos(\theta_{r,w})$ & \multicolumn{1}{|c|}{$-$} & $R\cos(\theta_{r,w})$ \\\cline{2-6}
\multicolumn{1}{|c|}{\phantom{ }} & $\mathrm{to}$ & $-$ & $\mathrm{PO}(r,w)$ & \multicolumn{1}{|c|}{$-$} & $\mathrm{PO}(r,w)$ \\\hline
\multicolumn{1}{c|}{\phantom{ }} & at $d=R$ & $\mathrm{PO}(r+1,w)$  & $\mathrm{PO}(r+1,w)$ & \multicolumn{1}{|c|}{$\mathrm{PO}(r-1,w)$} & $\mathrm{PO}(r+1,w)$\\\cline{2-6}
\end{tabular}\end{center}
\caption[Summary of the diffractive orbit bifurcations for the circle billiard.]{\label{tab:bifurcations}
Summary of the diffractive orbit (DO) bifurcations for the circle billiard, which are separated into three groups.
The table should be read columnwise and is explained in the text.
}
\end{table*}

Using the results in \cite{Schomerus1997} for the uniform approximations of periodic orbits involved in pitchfork and
tangent bifurcations, we can write down the corresponding formulae for the uniform approximation of pairs of diffractive
orbits involved in the bifurcations in Tab.~\ref{tab:bifurcations}. For our pitchfork bifurcations involving a diffractive orbit
labelled by $0$ (exists before and after the bifurcation) and a pair of periodic orbits labelled by $1$ (exist only after the
bifurcation), the uniform approximation is

\begin{eqnarray} \label{eq:uniform_pitchfork}
d_{\mathrm{PF}}(k) &=& \Real \left| 2 \pi k \Delta L \right|^{1/2} (-1)^{r_d} \exp \left[ \ui \left( k \bar{L} - \frac{\pi}{2} \bar{\nu}
- \frac{3 \pi}{4} \mu_d \right) \right] \nonumber \\
& \times & \Bigg[ \left( \frac{A_1}{2} + \frac{A_0}{\sqrt{2}} \right) \left( \sigma_2 J_{1/4} \left(k |\Delta L| \right) \expp^{\ui \sigma_1 \pi / 8}
+ J_{-1/4} \left(k |\Delta L| \right) \expp^{-\ui \sigma_1 \pi / 8} \right) \nonumber \\
&+& \left( \frac{A_1}{2} - \frac{A_0}{\sqrt{2}} \right) \left(J_{3/4} \left(k |\Delta L| \right) \expp^{\ui \sigma_1 3 \pi / 8}
+ \sigma_2 J_{-3/4} \left(k |\Delta L| \right) \expp^{-\ui \sigma_1 3 \pi / 8} \right) \Bigg] \,, \nonumber \\
\end{eqnarray}
where $\bar{L}=(L_1+L_0)/2$ and $\Delta L=(L_1 - L_0)/2$. The amplitudes $A_0$ and $A_1$ are given by (\ref{eq:amplitude}),
$\sigma_1 = \mathrm{sign}(\Delta L)$ and $\sigma_2 = -1$ before the bifurcation and $\sigma_2=1$ after the bifurcation.
The index $\bar{\nu}$ is the average Maslov index of the orbits after the bifurcation. In our case we always have pitchfork
bifurcations involving a diffractive orbit and a corresponding diffractive periodic orbit.

For the uniform approximation of a  tangent bifurcations, it is useful to complexify the phase space and view the birth of
two (real) periodic orbits as to originate from two complex periodic orbits (so callled \emph{ghost orbits}) prior to the bifurcation. 
Labeling the two real orbits by $1$ and $2$ gives the contribution from the uniform approximation after the bifurcation
\begin{eqnarray}\label{eq:uniform_tangent_1}
d_{\mathrm{TG}}(k) &=& \Real \left| \frac{8 \pi k \Delta L}{3} \right|^{1/2} (-1)^{r_d} \exp \left[ \ui \left(k \bar{L} - 
\frac{\pi}{2} \bar{\nu} - \frac{3 \pi}{4} \mu_d \right) \right] \nonumber \\
&\times& \Bigg[ \frac{A_1+A_2}{2} \left( J_{-1/3} \left(k |\Delta L| \right) + J_{1/3} \left(k |\Delta L| \right) \right) \nonumber \\
&-& \ui \mathrm{sign}(\Delta L) \, \frac{A_1 - A_2}{2} \left(J_{-2/3} \left(k |\Delta L| \right) - J_{2/3} \left(k |\Delta L| \right)
\right) \Bigg] \,, \nonumber \\
\end{eqnarray}
where $\bar{L}$ and $\bar{\nu}$ are the average length and Maslov index, respectively, and $\Delta L = L_1 - L_2$.
The ghost orbits before the bifurcation can be computed from the complex solutions of
(\ref{eq:difforb_cond_alpha}).  It is then convenient to write the contribution from the uniform approximation before the bifurcation in the form 

\begin{eqnarray} \label{eq:uniform_tangent_2}
d_{\mathrm{TG}}(k) &=& \Real \left| \frac{8 k \Delta L}{\pi} \right|^{1/2} (-1)^{r_d} 
\exp \left[\ui \left(k \bar{L} - \frac{\pi}{2} \bar{\nu} -\frac{3 \pi}{4} \mu_d \right) \right] \nonumber \\
&\times& \Bigg[ \frac{A_1+A_2}{2} K_{1/3} \left(k |\Delta L| \right) + \frac{A_1-A_2}{2} K_{2/3}
\left(k |\Delta L| \right) \Bigg] \,, \nonumber \\
\end{eqnarray}
with $\bar{L}$, $\Delta L$ and $\bar{\nu}$ as above and $K$ being the modified Bessel function of the second kind. In this case,
the lengths, actions and monodromy matrices are all complex valued.
For the correct way to complexify, see \cite{Schomerus1997}.

In some cases we need to consider also multiply diffractive orbits in which one singly diffractive part bifurcates.
In these cases, the uniform approximations above still hold when inserting the quantities of the full multiply
diffractive orbits.

As the scatterer is moved from the centre of the billiard toward the edge, the bifurcations happen in sequence and,
for a given value of $d$, one must determine which bifurcations are within its vicinity. 
In most cases this consists of determining which tangent bifurcation was last to occur and which is next to occur.
For a fixed value of $d$ we determine the next tangent bifurcation to occur in the direction of increasing $d$ and
input the corresponding ghost orbits into the diffractive orbit sum.

In the upper panel of Fig.~\ref{fig:length_spec_diff_closed} we show the length spectrum difference $\Delta F$ 
together with the short-wave approximation using the uniform approximations discussed above for $d=0.59$, $a=1$ and
$\mu_{\mathrm{max}}=4$. Our choice of scatterer position is generic as it is not exactly at a bifurcation point
and is between pitchfork and tangent bifurcations allowing us to apply both types of uniform approximation.
Comparing this to the lower panel we see that the uniform approximations have improved the correspondence.
In fact, we can now identify the discrepancies in the lower panel with particular orbits. We see discrepancies
near the lengths corresponding to the diffractive orbit $\mathrm{DO}(3,1)$ (of length $l=5.2204$), the 
diffractive periodic orbit $\mathrm{PO}(3,1)$ (of length $l=5.1962$) that was created from it in a 
pitchfork bifurcation at $d=0.5$, and the real part of the lengths of the ghost orbits $\mathrm{DO}(4,1)$
and $\mathrm{DO}(4,1)'$ ($l=5.7196$). For $d=0.5$ the scatterer would be positioned exactly at the
bifurcation point of the pitchfork bifurcation. In this instance one has to use the limiting form of the 
corresponding uniform approximation \cite{Schomerus1997}.

\begin{figure}
\centerline{
\includegraphics[angle=0,width=16cm]{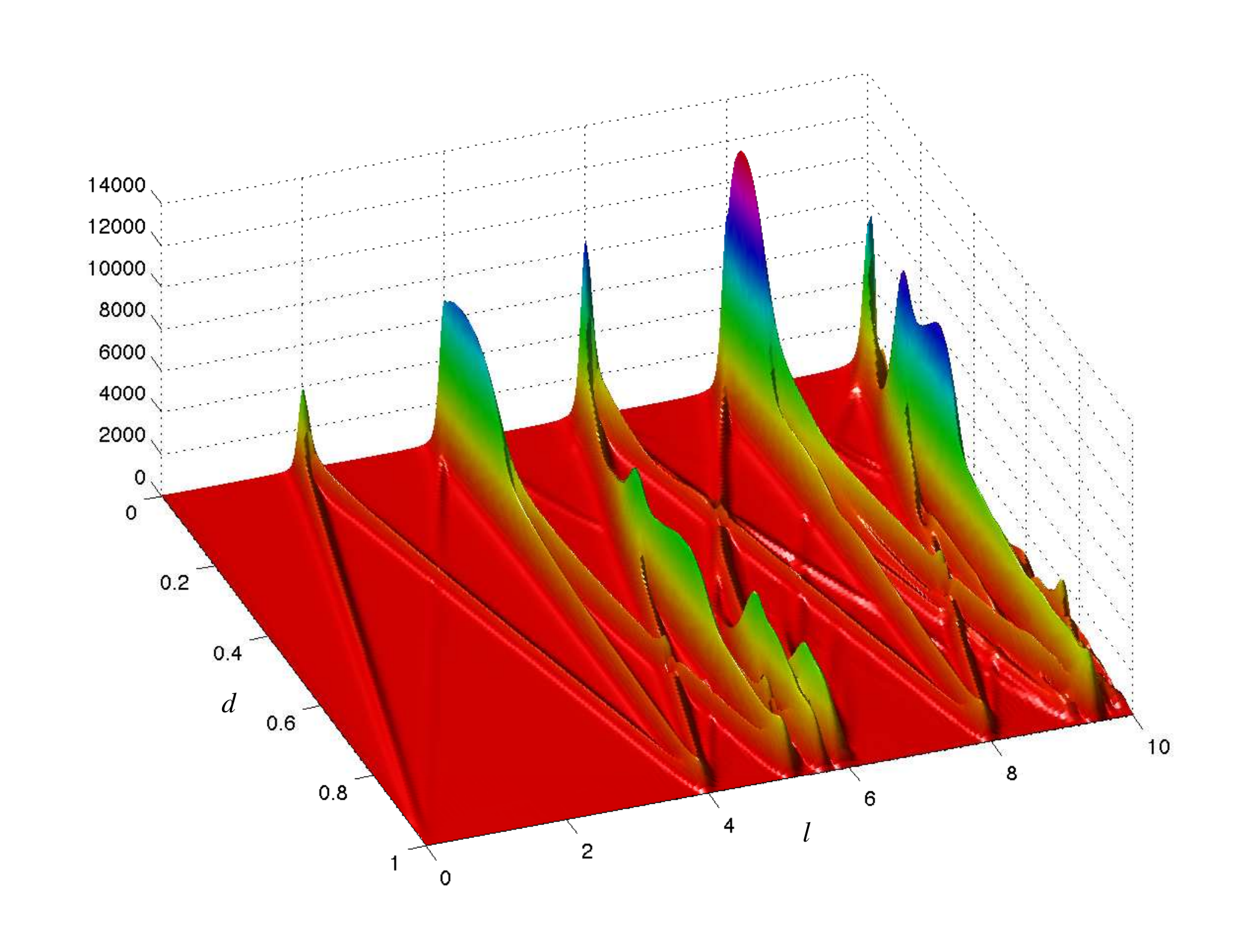}
}
\caption[Three-dimensional plot of the absolute value of the exact length spectrum.]{\label{fig:length_spec_3d}
Three dimensional plot of the absolute value of the exact length spectrum difference $\vert \Delta F(l) \vert$
for the perturbed circular billiard for varying position $d$ of the scatterer, as computed from
the resonance spectra. ($a=1.0$, $t=0.001$.)
}
\end{figure}

\begin{figure}
\centerline{
\includegraphics[angle=0,width=16cm]{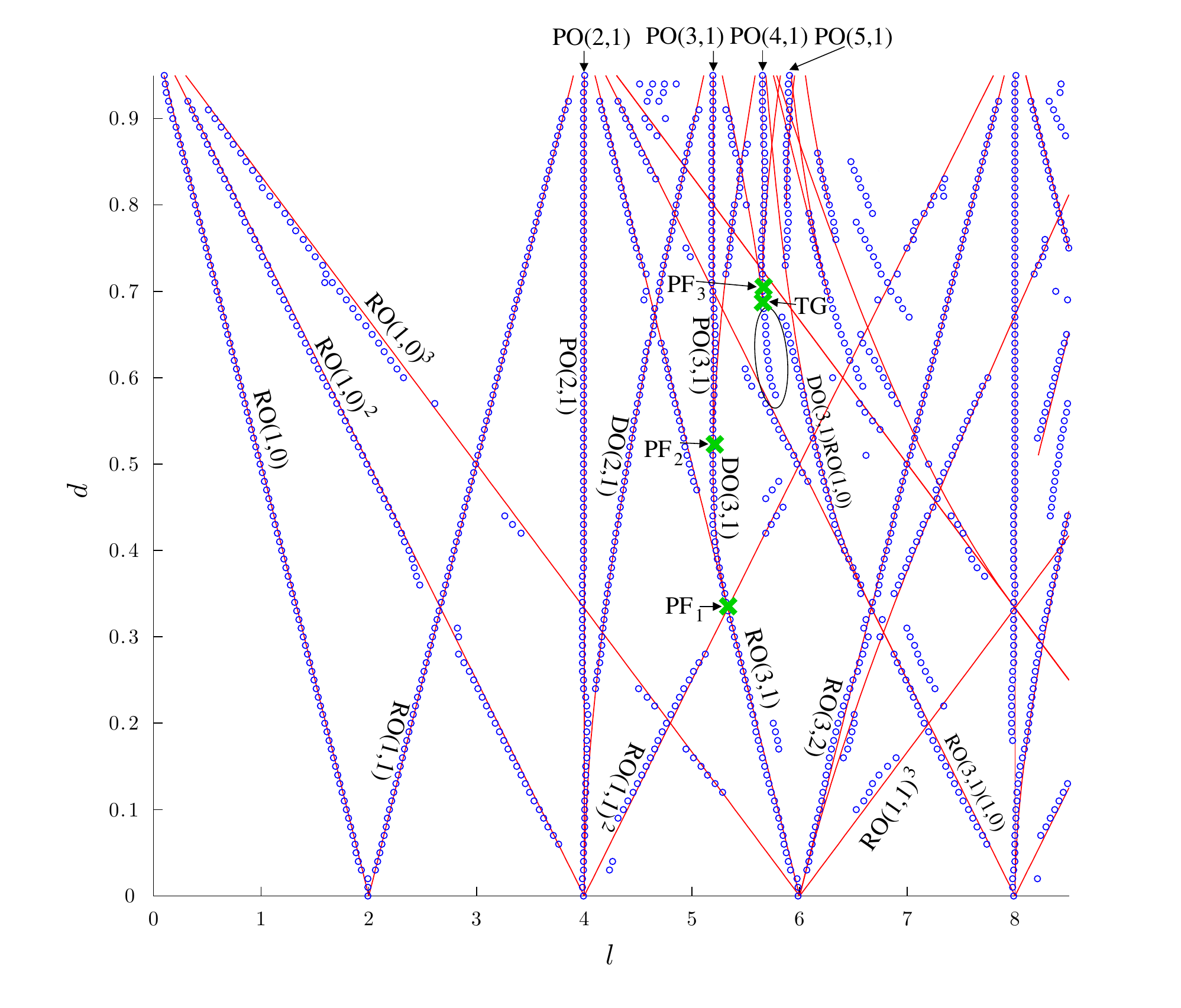}
}
\caption[Bifurcations in the circle billiard with point scatterer.]{\label{fig:bifurcation_plot}
Bifurcations in the circle billiard with point scatterer. The circles correspond to the maxima of the exact length
spectrum difference in Fig.~\ref{fig:length_spec_3d} and the lines are the lengths of the classical diffractive orbits.
They are labelled as described in the text. Two bifurcation sequences are highlighted. The first sequence 
involves two pitchfork bifurcation: first the orbit $\mathrm{DO}(3,1)$ splits from $\mathrm{RO}(3,1)$ at $\mathrm{PF}_1$,
and later $\mathrm{PO}(3,1)$ splits from $\mathrm{DO}(3,1)$ at $\mathrm{PF}_2$. In the second sequence
$\mathrm{DO}(4,1)$ and $\mathrm{DO}(4,1)'$ are created in a tangent bifurcation at $\mathrm{TG}$,
and then $\mathrm{PO}(4,1)$ splits from $\mathrm{DO}(4,1)$ in the pitchfork bifurcation $\mathrm{PF}_3$.
Before the tangent bifurcation we circle maxima of the length spectrum which correspond to ghost orbits.
Note that for illustrative purposes not all classical orbits are marked in the plot.
}
\end{figure}

In Fig.~\ref{fig:length_spec_3d} we give a three-dimensional plot of the exact length spectrum difference $\Delta F(l)$
for the scatterer varying from the centre of the billiard to the boundary, i.e. $d \in [0,1]$. 
In Fig.~\ref{fig:bifurcation_plot} we plot the maxima of this plot in the $(d,l)$-plane together with the lengths of the
diffractive orbits. We are able to identify all of the expected diffractive orbits. In particular,
we can identify multiple diffractions of the shortest radial diffractive orbit $\mathrm{RO}(1,0)$ and combinations
of longer orbits with this radial orbit. We can also identify the results of broken tori at $d=0$ and sequences
of two pitchfork bifurcations whence the diffractive orbits $\mathrm{DO}(2k+1,k)$ with $k \in \mathbb{N}_0$ are
created in the first bifurcation and the associated periodic orbits $\mathrm{PO}(2k+1,k)$ are created in the second.
Further, we can identify the sequences of tangent and pitchfork bifurcations whence the pair of diffractive orbits
$\mathrm{DO}(r,w)$ and $\mathrm{DO}(r,w)'$ are created in the tangent bifurcation and the associated periodic
orbit $\mathrm{PO}(r,w)$ is then created from $\mathrm{DO}(r,w)'$ in the pitchfork bifurcation. Additionally, for
the tangent bifurcations, we see a  trail of maxima leading from the bifurcation point that we identify with the
ghost orbits included in the uniform approximation (\ref{eq:uniform_tangent_2}).

In the TG-PF bifurcation sequences, the two bifurcations always lie very near to each other with respect to the
bifurcation parameter $d$, which causes a problem when implementing the two corresponding uniform approximations.
This has not yet been addressed and is remarked upon in \cite{Brack2009}. As explained above, for a generic scatterer
position we always consider the nearest bifurcation(s). For the value $d=0.59$ used in the figures,
this is pitchfork bifurcation at $d=0.5$ and the tangent bifurcation at $d=0.6825$.

\section{The open dielectric cavity} \label{sec:unptbd_open}

Having studied the closed cavity we now turn to the optical microcavity, which is a circular disk
of dielectric material of refractive index $n>1$ in an outside medium of lower refractive index $\tilde{n}=1$.
Waves can now reflect and refract at the interface and, restricting ourselves to TM polarisation,
we consider the reduced Maxwell equations in the form of the Helmholtz equation (\ref{eq:hholtz_2d}). 

\subsection{Exact solutions} \label{sec:unptbd_open_exact}

The spectrum for outgoing boundary conditions consists of complex-valued resonances of the form
$k=k_r+\ui k_i$, with $k_i<0$. It can be found from the Green function. Inside the cavity,
the outgoing Green function is given in polar coordinates $\mathbf{r}=(r,\phi)$ by
\cite{Dettmann2008,Dettmann2009uni}
\begin{equation}\label{eq:green_func1}
G(\mathbf{r},\mathbf{r}',k) = -\frac{\ui}{4} H_0(kn\vert\mathbf{r}-\mathbf{r}'\vert) + 
\frac{\ui}{4} \sum_{m=-\infty}^{\infty} \frac{C_m}{A_m} \cos\left[m(\phi-\phi')\right] J_m(knr) J_m(knr')\,.
\end{equation}
The coefficients in this expansion are given by
\begin{eqnarray}\label{eq:green_coeffs1}
A_m &=& J_m(knR) H'_m(kR)  - n J'_m(knR) H_m(kR) \,, \nonumber \\
C_m &=& H_m(knR) H'_m(kR) - n H'_m(knR) H_m(kR) \,,
\end{eqnarray}
which, by using the formula for Bessel function derivatives \cite{AbramSteg1964}, we write as
\begin{eqnarray}\label{eq:green_coeffs2}
A_m &=& n H_m(kR) J_{m+1}(knR)  - J_m(knR) H_{m+1}(kR) \,, \nonumber \\
C_m &=& n H_m(kR) H_{m+1}(knR) - H_m(knR) H_{m+1}(kR) \,,
\end{eqnarray}
which is advantageous for numerical purposes. The spectrum  is obtained from the poles of (\ref{eq:green_func1}) which are given by
$A_m=0$ and hence each resonance is determined as a complex root of the
function \cite{Dettmann2008}
\begin{equation}\label{eq:quantcond_unpted_1}
f_m(k) = n H_m(kR) J_{m+1}(knR) - J_m(knR) H_{m+1}(kR) \,.
\end{equation}
The corresponding eigenfunctions for a solution of $f_m(k)=0$, which represent the $z$-component of the
electric field, have the form
\begin{equation}\label{eq:solution_open_in}
\psi(r,\phi) =\frac{H_m(kR) J_m(knr)}{J_m(knR)}
\begin{cases}
\sin(m\phi)\,, & -m\in\mathbb{N} \\
\cos(m\phi)\,, & m\in\mathbb{N}_0
\end{cases} \,,
\end{equation}
for $r<R$ and 
\begin{equation}\label{eq:solution_open_out}
\psi(r,\phi)=H_m(kr)
\begin{cases}
\sin(m\phi)\,, & -m\in\mathbb{N} \\
\cos(m\phi)\,, & m\in\mathbb{N}_0
\end{cases} \,,
\end{equation}
for $r>R$.

We mentioned in Section~\ref{sec:trace} that there are two kinds of resonances, inner and outer. One way to 
distinguish between them is to consider the limit as the refractive index $n \rightarrow \infty$. In this limit
the system becomes closed and separates into an inside and an outside problem. The outer resonances
are those that correspond in this limit to solutions of the outside scattering problem with Dirichlet boundary
conditions at the disk \cite{Dubertrand2008,Bogomolny2008,Dettmann2009}. We will not consider them
further, because they lie deeper in the complex plane than the inner resonances, and they do not play a
role in the trace formula. The inner resonances on the other hand have the property that their wavenumbers
become real in the limit $n \rightarrow \infty$. 
We label these resonances by two indices $m$ and $q$ where for fixed $m$, the index $q$ enumerates the resonances in the  order of increasing real part of their wavenumbers
starting from $q=1$.  
By considering the asymptotic behaviour of the function
$f_m(k)$ in (\ref{eq:quantcond_unpted_1}) one finds that the wavenumbers of the resonances labeled this way have the limiting behaviors 
\cite{Ryu08,Dettmann2009}
\begin{align} \label{eqn:smallopenlim}
\lim_{n \rightarrow \infty} n k_{m,q} R &= j_{m-1,q}, \quad m \neq 0 \nonumber \\
\lim_{n \rightarrow \infty} n k_{0,q}  R &= j_{1,q-1}, \quad q \neq 1 \quad \\
\lim_{n \rightarrow \infty} n k_{0,1}  R &= 0. \nonumber
\end{align}
On the one hand this limiting behaviour can conversely be used in order to label the resonances uniquely
by two indices $m$ and $q$, where $m \in \mathbb{Z}$ and $q \in \mathbb{N}$. On the other
hand the relations (\ref{eqn:smallopenlim}) are very helpful for finding the resonances numerically. 
In general, it is quite difficult to systematically find all solutions of the equation $f_m(k)=0$ for
finite $n$ in some region of the complex plane. For this reason, we start with the resonances
for some high value of the refractive index $n$ for which the values in (\ref{eqn:smallopenlim})
serve as good approximations for the wavenumbers. We found it convenient to start with 
$n = 50$. We then gradually decrease $n$ and follow the resonances in the complex plane.
We note that the standard routines for computing the Bessel and Hankel functions with complex
arguments (e.g. NAG, Maple) were not sufficient for computing many of those resonances which lie
very close to the real axis. We computed the results here using the routine of Amos (\cite{Amos1983-1},
\cite{Amos1983-2}) with the modifications given in \cite{Yousif1997} and \cite{Yousif2003}. This is a very
sensitive computation for high modal numbers $m$, and in some instances will have false convergences
to numbers with very small but positive imaginary parts. This happens when the actual solutions are 
very close to the real axis, and we set the imaginary parts to be numerically zero in these cases.
Further, we found that the starting value $n_{\mathrm{max}}$ should be decreased with increasing $m$
so that by $m=50$ we took $n_{\mathrm{max}}\approx7$.

\begin{figure}
\centerline{\includegraphics[angle=0,width=12cm]{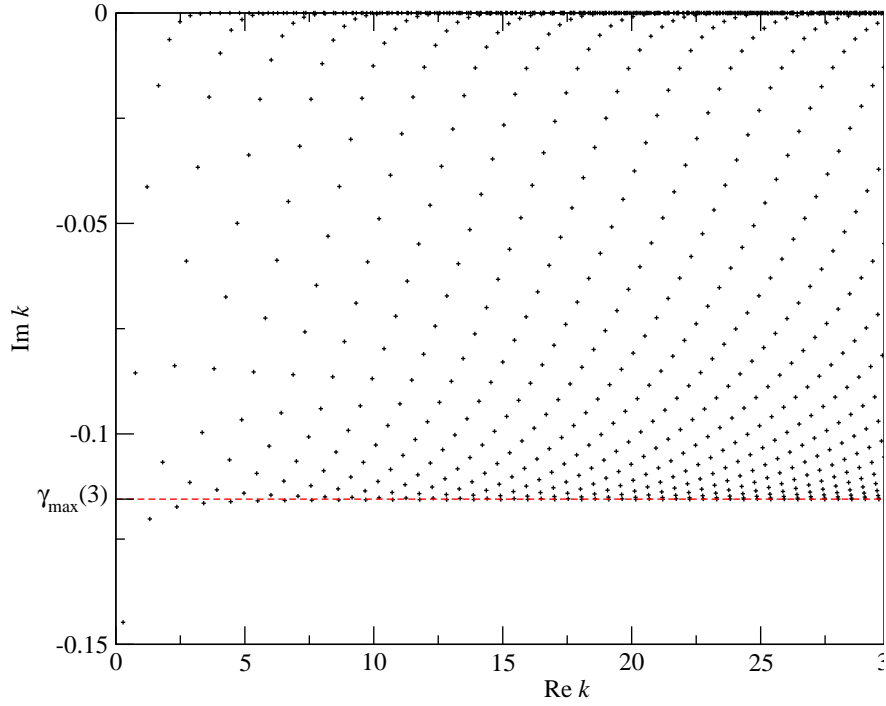}}
\caption[Resonances in the complex wavenumber plane for two refractive indices.]{\label{fig:res_unptbd}
Resonances in the complex plane of the unperturbed circular microcavity with refractive index $n=3$. Each resonance
string corresponds to a fixed radial modal number $q$ and the modal number $m$ increases as one moves along the
string from the threshold $\gamma(n)$ (dashed line) towards the real axis.}
\end{figure}

By the symmetry of the Bessel functions we have that $k_{m,q}=k_{-m,q}$ and hence all resonance
wavenumbers with $m \neq 0$ are double degenerate. We denote the multiset of the wavenumbers
of the inner resonances again by $\sigma$. Then
\begin{equation}
\sigma = \{k_{m,q} \in \mathbb{C}: m \in \mathbb{Z}, q \in \mathbb{N}, f_m(k_{m,q}) = 0 \} \,.
\end{equation}
In Fig.~\ref{fig:res_unptbd} we show the resonances for $n=3$. One can see that, except for a few
of the first resonances, they all lie in an infinite strip bounded by the line \cite{Dubertrand2008}
\begin{equation} \label{eq:inner_thres_eq}
-\Imag \, k R \leq \gamma(n),
\end{equation}
where
\begin{equation}\label{eq:inner_thres}
\gamma(n) = \frac{1}{2n} \log \frac{n+1}{n-1} \,,
\end{equation}
This bound arises from the asymptotic forms of the Bessel and Hankel functions in (\ref{eq:quantcond_unpted_1})
for large $k$ and $m=0$.

The length spectrum is again given by the Fourier transform $F(l)$ of the density of states multiplied with the Gaussian
cut-off function $W(k)=\exp(-t n^2 k^2)$. We determined the resonances in a window $0 < \Real \, k < k_{\mathrm{max}}$ 
with $k_{\mathrm{max}}=100/n$ and chose the cut-off parameter $t=10/(n k_{\mathrm{max}})^2$. We hence define
\begin{equation}\label{eq:trace_qm_fourier}
F(l) = \int_0^\infty d(k) \, W(k) \expp^{- \ui n k l} \ud k \approx \sum_{m,q} \expp^{-\ui n k_{m,q} l - n^2 k_{m,q}^2 t} \,.
\end{equation}
In contrast to the closed case, the density of states is now given by a sum over Lorentzians, see Eq.~(\ref{eq:dstate_res}).
The integral can be expressed in terms of the error function with complex argument, and the approximation in
equation (\ref{eq:trace_qm_fourier}) holds in our parameter range.

\subsection{Short-wave approximation} \label{sec:unptbd_open_sc}

The short-wave approximation for the length spectrum is obtained by inserting the trace formula
for the density of states into the definition (\ref{eq:trace_qm_fourier}). In section \ref{sec:trace}
it was discussed how the trace formula is modified for the open cavity in comparison to the closed 
cavity. In the smooth part of the density of states only the perimeter term changes and we obtain
\begin{equation}\label{eq:trace_smooth_fourier}
\Real F_{0}(l) = \frac{\mathcal{A}}{4 \pi t} + \frac{\expp^{-l^2 / 4 t}}{8 \sqrt{\pi t}}
\left[\frac{\tilde{r}(n) L}{n} + \frac{\ui \mathcal{A} l}{t} \mathrm{erf} \left(\frac{\ui l}{2 \sqrt{t}} \right) \right] \,,
\end{equation}
where $\mathcal{A}=\pi R^2$, $L=2 \pi R$ and $\tilde{r}(n)$ is defined in (\ref{eq:r_tilde}).

The oscillatory part is changed according to equation (\ref{eq:d_xi2}): One has an additional overall
factor of $n$ and the wavenumber $k$ is replaced by $n k$. Furthermore, the factors that are due
to the boundary condition at the cavity wall change. Instead of a Dirichlet phase factor $(-1)$ for every
reflection we now have to include a Fresnel coefficient for every reflection. For TM modes it is given by
\begin{equation} \label{eq:reflec_coeff}
R(\theta) = \begin{cases}
\frac{n \cos \theta - \sqrt{1 - n^2 \sin^2 \theta}}{n \cos \theta + \sqrt{1 - n^2 \sin^2 \theta}} \,,
& \theta < \theta_c(n) \, , \\
\frac{n \cos \theta - \ui \sqrt{n^2 \sin^2 \theta - 1}}{n \cos \theta + \ui \sqrt{n^2 \sin^2 \theta - 1}} \,,
& \theta > \theta_c(n) \, ,
\end{cases}
\end{equation}
where $\theta$ is the angle of incidence and the critical angle $\theta_c$ is defined in (\ref{eq:crit_angle}).
This coefficient allows for leakage from the cavity and/or entrapment within it (total internal reflection).
For trajectories in the circular cavity the angle of incidence $\theta$ is the same for each of its reflections,
and hence we find that the oscillatory part for the closed cavity in (\ref{eq:trace_osc_closed}) changes into
\begin{equation}\label{eq:trace_osc}
d_{\mathrm{po}}(k) = \frac{n}{\pi} \sum_{r,w} g_p \mathcal{A}_p \sqrt{\frac{2 n k}{\pi l_p}} \, R_p^r
\expp^{\ui n k l_p - \ui r \pi / 2 + \ui \pi / 4}  + \mathrm{c.c.} \,.
\end{equation}
for the open cavity. The index $p=(r,w)$ again labels the periodic orbit families with $w$ rotations around
the billiard centre and $r$ reflections at the boundary. The quantities in this formula have been discussed
after the trace formula for the closed cavity (\ref{eq:trace_osc_closed}), except for the Fresnel reflection
coefficient $R_p$ which is evaluated at the angle of incidence (\ref{eq:ang_inci_po}) of each periodic orbit.

We insert the periodic orbit contributions (\ref{eq:trace_osc}) into the definition of the length spectrum
and perform the integral analytically. Again we consider only the real part and obtain
\begin{align} \label{eq:trace_osc_fourier}
\Real F_{\mathrm{po}}(l) & = \frac{1}{2 \pi} \sum_p
\frac{g_p \, \mathcal{A}_p \, |R_p|^r \, (-1)^{\lfloor (r+1)/2 \rfloor}}{ \sqrt{2 l_p}} \, \left( \frac{2}{t} \right)^{3/4} \times
\\ \notag & \left\{ \expp^{-(l_p+l)^2/8 t} \left[ 
\sin \left( \frac{\pi r}{2} - r \phi_p \right) D_{1/2} \left( \frac{(l_p+l)}{\sqrt{2 t}} \right) + 
\cos \left( \frac{\pi r}{2} - r \phi_p \right) D_{1/2} \left( -\frac{(l_p+l)}{\sqrt{2 t}} \right)
\right] \right.
\\ \notag & \left.  + \expp^{-(l_p-l)^2/8 t} \left[ 
\sin \left( \frac{\pi r}{2} - r \phi_p \right) D_{1/2} \left( \frac{(l_p-l)}{\sqrt{2 t}} \right) +
\cos \left( \frac{\pi r}{2} - r \phi_p \right) D_{1/2} \left( -\frac{(l_p-l)}{\sqrt{2 t}} \right) 
\right] \right\} \, .
\end{align}

\begin{figure}
\centerline{\includegraphics[angle=0,width=7cm]{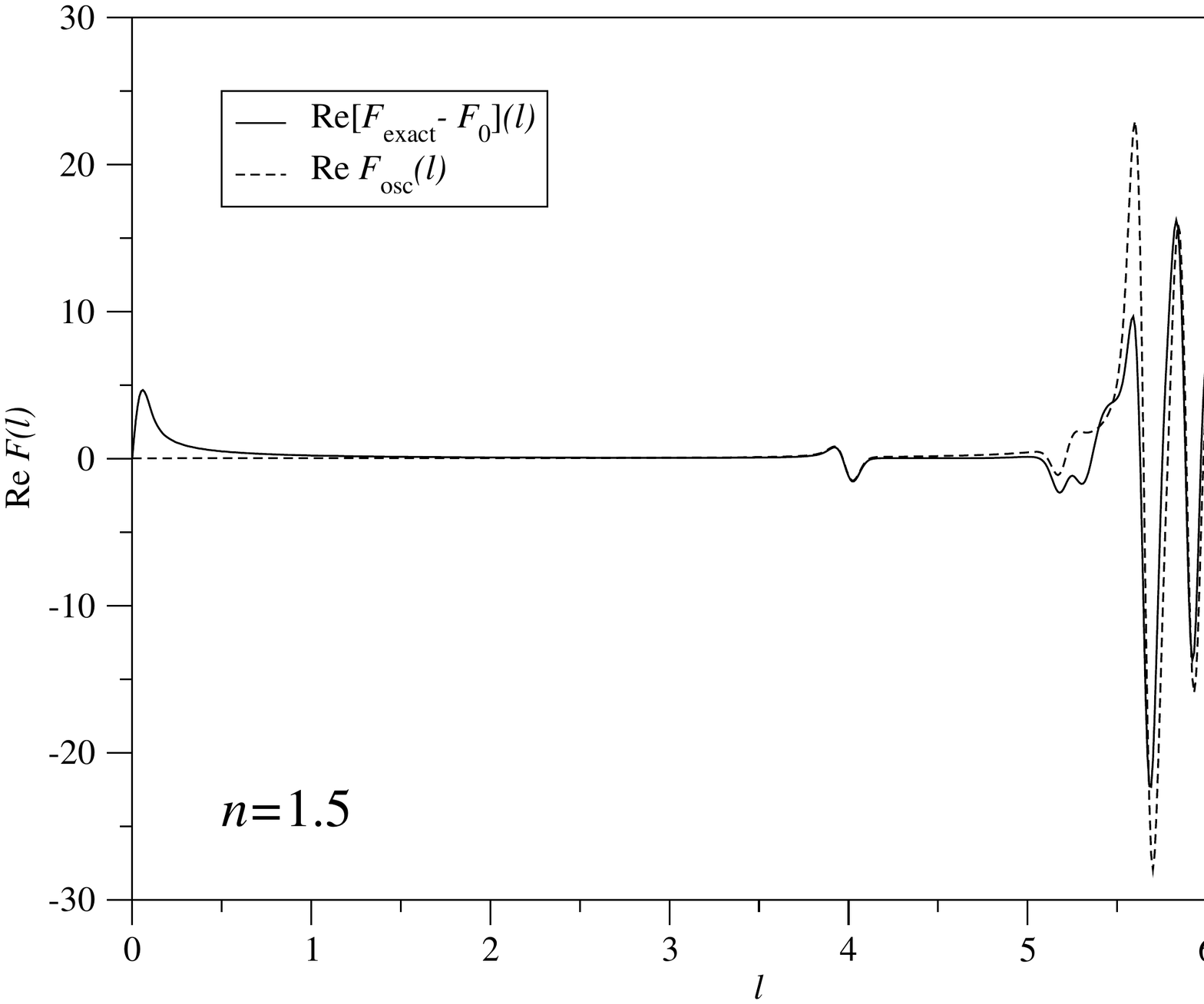} \quad \includegraphics[angle=0,width=7cm]{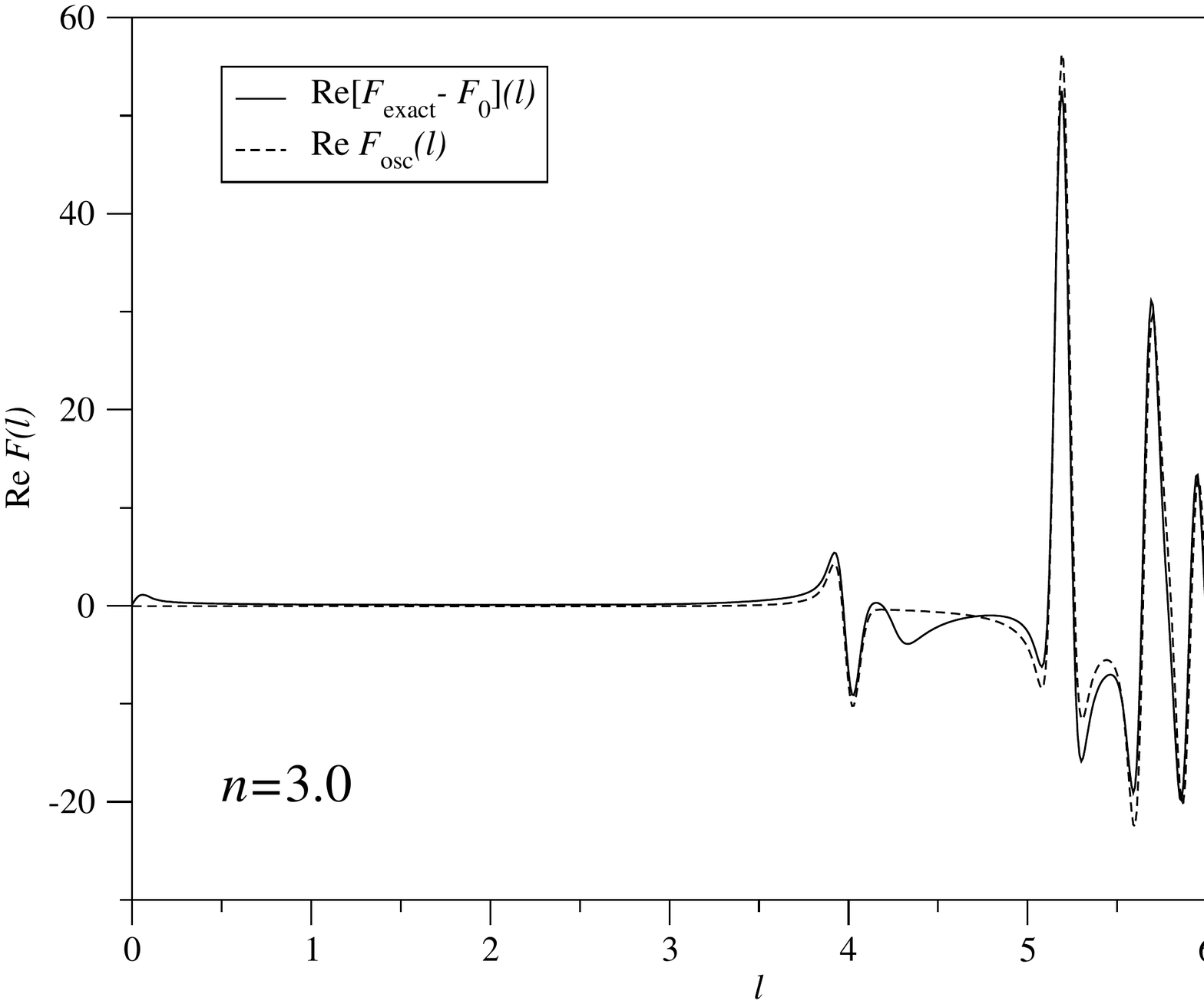}}
\vspace{1cm}
\centerline{\includegraphics[angle=0,width=7cm]{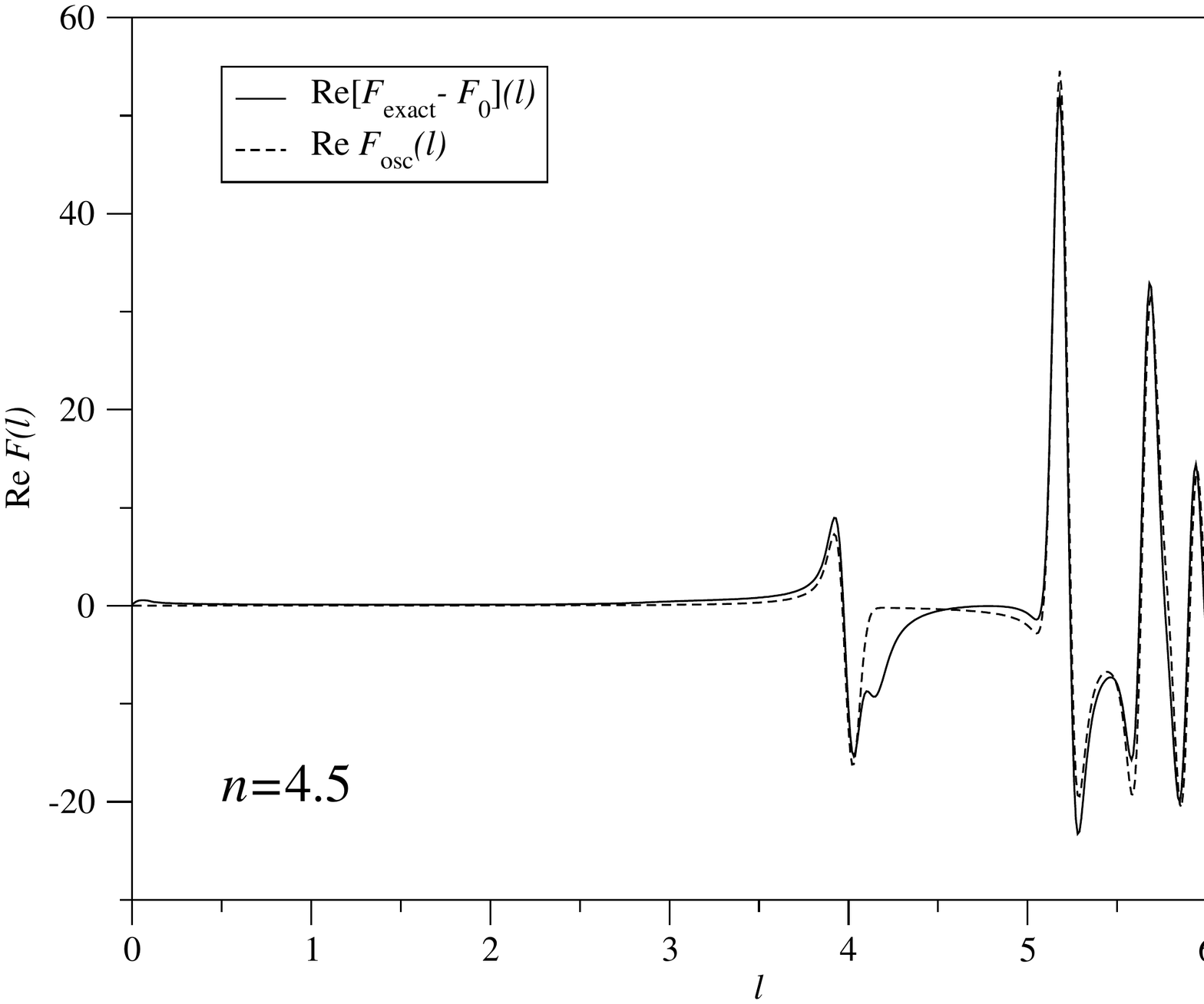}}
\caption[Exact and semiclassical length spectra for three refractive indices.]{\label{fig:length_spec_open}
Length spectra for the circular dielectric cavity, for three values of the refractive index $n$. The exact spectrum
is computed from (\ref{eq:trace_qm_fourier}) with unperturbed resonances and the short-wave approximation
from (\ref{eq:trace_osc_fourier}) with periodic orbits. ($t=0.001$.)
}
\end{figure}

\begin{figure}
\centerline{
\includegraphics[angle=0,width=10cm]{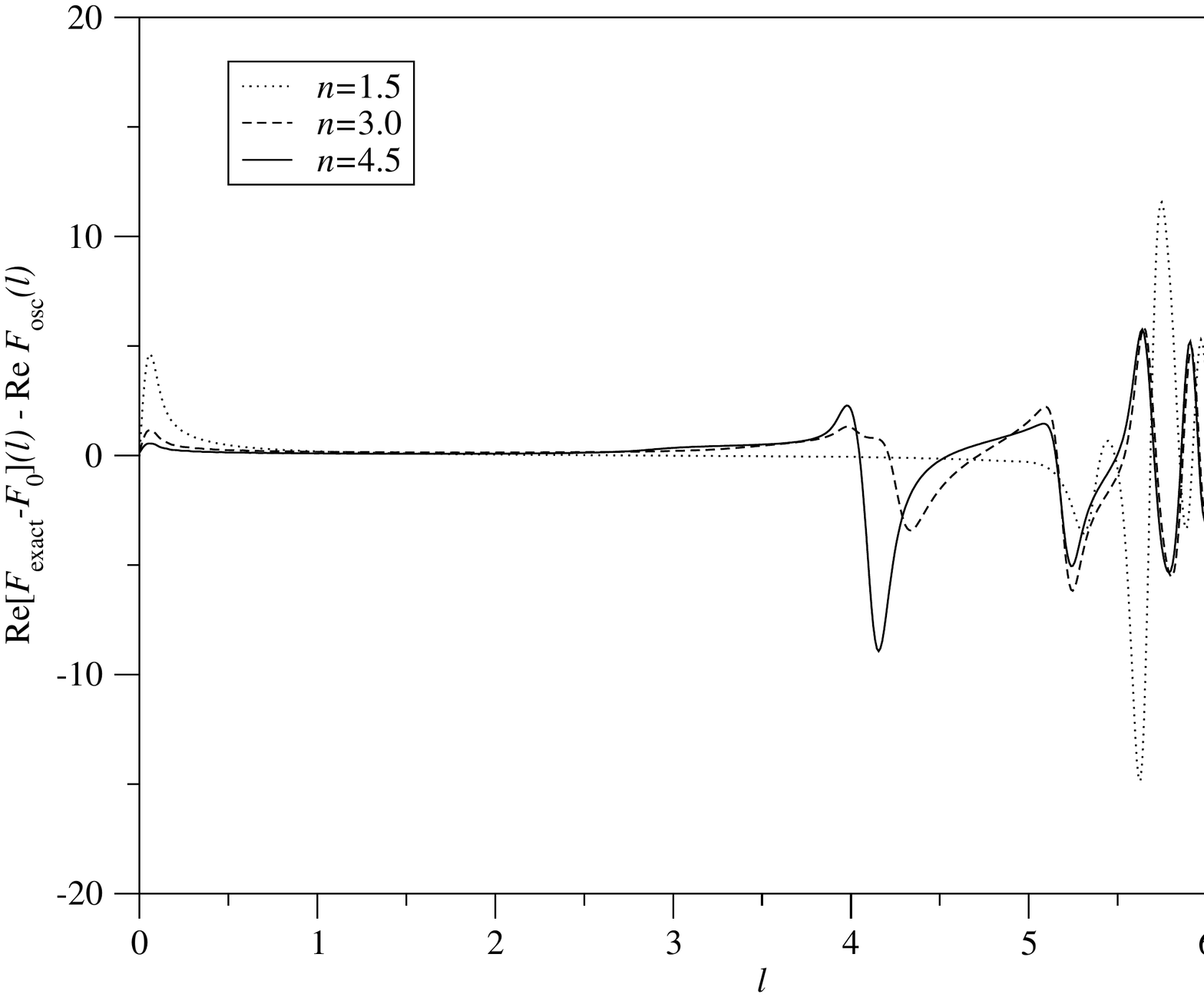}
}
\caption[Difference between the exact and semiclassical length spectra.]{\label{fig:length_spec_open_sc_diff}
Difference between the exact length spectrum of the circular dielectric cavity and its short-wave approximation
for the three values of $n$ in Fig.~\ref{fig:length_spec_open}. ($t=0.001$.)
}
\end{figure}

As mentioned before, the formula becomes less accurate when the angles of incidence of the contributing
periodic orbits are close to the critical angle \cite{Hentschel2002}. We can therefore estimate where we
expect an error to be located in the length spectrum. If we insert the critical angle into the formula for
the periodic orbit lengths (\ref{eq:po_length}) and use the interpolating formula for the number of reflections
$r=(\pi - 2 \theta)/2 \pi$ of orbits with $w=1$ rotations we obtain
\begin{equation}\label{eq:po_crit_angle}
l_p(n) = \frac{4 \pi R \sqrt{n^2-1}}{n \left(\pi - 2 \theta_c(n) \right)} \,.
\end{equation}
We have that $l_p(n) \rightarrow 2 \pi R$ as $n \rightarrow 1_+$ and $l_p(n) \rightarrow 4 R$ as $n\rightarrow\infty$,
which correspond to the perimeter and twice the diameter of the circle, respectively. We therefore have that for suitably
large refractive index the error is not located near the accumulation point $l=2 \pi R$ of the periodic orbits but near
$l=4R$ at which length we have only the diameter orbit $\mathrm{PO}(2,1)$.

In Fig.~\ref{fig:length_spec_open} we plot the oscillating part of the semiclassical length spectrum $\Real F_{\mathrm{po}}(l)$,
together with the exact length spectrum minus the smooth part $\Real[F(l)-F_{\mathrm{0}}(l)]$ for three different values of 
the refractive index. Although the periodic orbit structures are well reproduced by the short-wave approximation the error
is clearly bigger than for the closed cavity, in particular for the case $n=1.5$. In order to see the error more clearly, we
plot in Fig.~\ref{fig:length_spec_open_sc_diff} the difference between the exact and semiclassical length spectra. 
The lengths which correspond to the critical angle are $l=5.57$ for $n=1.5$, $l=4.81$ for $n=3$, and $l=4.55$ for $n=4.5$,
and we can indeed observe large errors at these positions. There are observable errors at other lengths, but they 
correspond to smaller relative errors, because they occur at peaks in the length spectrum in Fig.~\ref{fig:length_spec_open}.
They might be related to the Goos-H\"anchen shift \cite{Tureci2002} whereby the observed length at which one has peaks
in the exact spectrum is slightly shifted from the length at which one has the periodic orbits. We also see an error near $l=0$
in the length spectrum which decreases with increasing $n$. This results from only having two terms in the smooth part
(\ref{eq:trace_smooth}).

Our test of the trace formula for the dielectric circle goes beyond the results in \cite{Bogomolny2008} in that we checked
not only the positions of the peaks in the length spectrum but also their shapes.
Figs.~\ref{fig:length_spec_open} and \ref{fig:length_spec_open_sc_diff} provide encouraging support for the use of short-wave
approximations in open dielectric systems and show that the trace formula works as well as can reasonably be expected.

\section{The dielectric cavity with point scatterer} \label{sec:ptbd_open}

\subsection{Exact solution}\label{sec:ptbd_open_exact}

As with the closed system we perturb the dielectric cavity by placing a point scatterer at a position $\mathbf{d}=(d,0)$
in Cartesian coordinates, see Fig.~\ref{fig:circle_cavity}.
The exact resonance spectrum for this perturbed open system
is again determined by equation (\ref{eq:saet_cond}) where we now have to include the regularized Green function of the
unperturbed dielectric disk. The regularization of the Green function is done exactly as before, by subtracting the
logarithmic divergence of the Hankel function $H_0$. The only difference is that its argument contains now $n k$
instead of $k$. We hence obtain that the perturbed resonances, $\tilde{k}$, are determined as the solutions of
$\tilde{f}(k)=0$, where $\tilde{f}$ is the transcendental function \cite{Dettmann2008}
\begin{equation}\label{eq:quantcond_pted}
\tilde{f}(k) = -\frac{\ui}{4} + \frac{1}{2\pi} \left(\log \frac{kna}{2} + \gamma \right) +
\frac{\ui}{4} \sum_{m=0}^{\infty} \frac{C_m}{A_m} \epsilon_m J^2_m(k n d)\,.
\end{equation}
As before, the parameter $a$ is a measure of the strength of the perturbation and replaces the coupling constant
$\lambda$ appearing in (\ref{eq:saet_cond}). The coefficients $A_m$ and $C_m$ are given in (\ref{eq:green_coeffs1})
and (\ref{eq:green_coeffs2}), and $\epsilon_m$ in (\ref{eq:epsilon_degen}).

As mentioned in Section~\ref{sec:ptbd_closed} one can realize the point scatterer experimentally for small values of $a$
by a small disk of radius $a$ with Dirichlet boundary conditions. As discussed in \cite{Dettmann2009uni} one can also
use a small hole of radius $b$, filled with dielectric material of refractive index $n_b$. These parameters are related to $a$ by
\begin{equation} \label{eq:finite_scatt}
\log \frac{n k a}{2} + \gamma \approx \frac{2}{b^2 k^2 (n_b^2 - n^2)} \,.
\end{equation}
The value of $a$ again ranges from $0$ to $\infty$, and  at these two limiting values the zeros of (\ref{eq:quantcond_pted})
correspond to the unperturbed resonances of Section~\ref{sec:unptbd_open_exact}. For finite values of $a$, one
finds again that the rank one perturbation changes only one of each pair of degenerate resonances for $m \neq 0$
of the unperturbed system. This is again the case because the wavefunctions with angular dependence $\sin(m\phi)$ in
(\ref{eq:solution_open_in}) have a nodal line on the $x$-axis and hence they and their corresponding wavenumber
are both unaffected by the presence of the scatterer. The other resonances have
wavefunctions with angular dependence $\cos(m\phi)$ in the unperturbed system, and as $a$ changes from $0$
to $\infty$ their wavenumbers move from one unperturbed value $k_{m,q}$ to another. We call these the perturbed
resonances. The full spectrum $\tilde{\sigma}$ of the perturbed system is therefore
\begin{equation} \label{eq:ptbd_res_open}
\tilde{\sigma} = \{\tilde{k}_{m,q} = k_{m,q} \in \sigma: -m, q \in \mathbb{N} \} \cup \{\tilde{k}_{m,q} \in \mathbb{C}:
m \in \mathbb{N}_0, q \in \mathbb{N}, \tilde{f}(\tilde{k}_{m,q})=0 \} \, .
\end{equation}
For the labelling we adopt again the rule that $\tilde{k}_{m,q}$ has the same $m$ and $q$ values as the unperturbed
resonance at which it arrives at $a=\infty$. If we would use the unperturbed resonance at $a=0$ for the labelling
we would again miss the first perturbed resonance. This resonance moves from  $k=k_{0,1} \in \sigma$ at $a=\infty$
towards the imaginary axis and then down along it as $a$ is decreased.

The computation of the perturbed resonances is a more delicate procedure than the computation of the unperturbed
resonances in the previous section. In order to find the perturbed resonances for a non-zero and finite value of $a_0$
we first set $a$ to zero and use an unperturbed resonance as an initial starting value in a Newton procedure. (For the
first resonance we start at $a=\infty$.) Note that we cannot use a bisection method for the open system. 
Because of the logarithmic dependence on $a$ in (\ref{eq:quantcond_pted}), we then vary $a$ exponentially towards
the desired value $a=a_{0}$. The function $\tilde{f}(k)$ has a pole at each unperturbed resonance $k=k_{m,q} \in \sigma$
because the corresponding $A_m$ in the denominator of its defining sum  (\ref{eq:quantcond_pted})  vanishes
at each $k=k_{m,q}$. To avoid this numerical problem we scale the function $\tilde{f}(k)$ by the particular $A_m$
when computing perturbed resonances for the set $\tilde{\sigma}$. In fact, because $a=0$ corresponds to an unperturbed
resonance, we start the numerical procedure at a small value $a=10^{-30}$ and then use an approximate solution
to (\ref{eq:quantcond_pted}) obtained from a perturbative approach. For small (or large) values of $a$ this gives the
value of the perturbed resonance $\tilde{k}_{m,q}$ as \cite{Dettmann2009uni}
\begin{equation} \label{eq:small_perturb}
\tilde{k}_{m,q} \approx k_{m,q} - \ui \pi \frac{C_m(k_{m,q})}{D_m(k_{m,q})} \frac{\epsilon_m J^2_m(k_{m,q} n d)}{2 R \log a } \,,
\end{equation}
where
\begin{equation}\label{eq:small_perturb_coeff}
D_m(k) = (n^2-1) J_m(n k R) H_m(k R)- \frac{A_m(k)}{k R}\,.
\end{equation}

\begin{figure}
\centerline{
\includegraphics[angle=0,width=12cm]{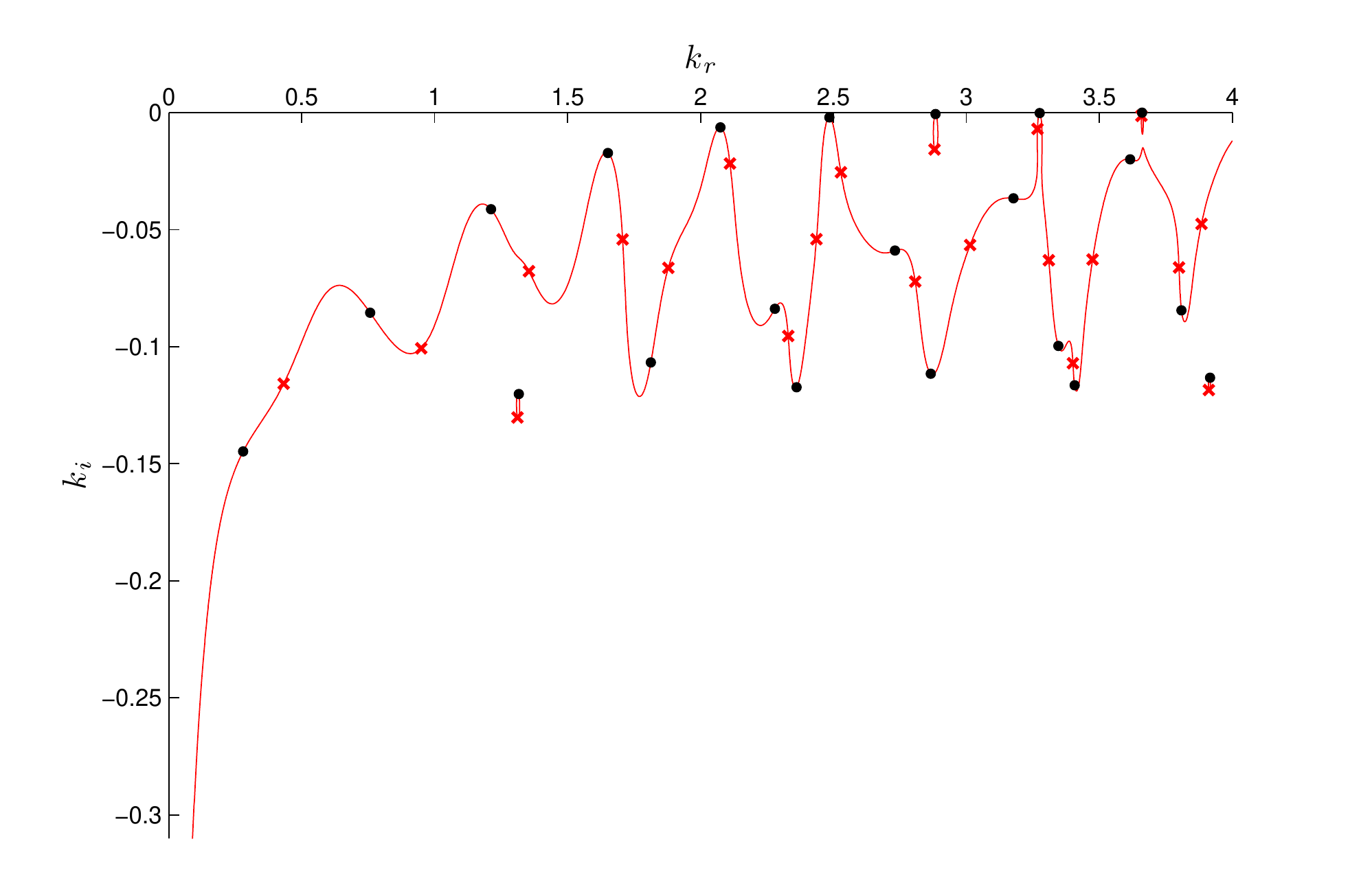}
}
\caption[Level dynamics of the perturbed resonances for $\Real k=0...4$.]{\label{fig:leveldyn_lowk}
Level dynamics of the perturbed resonances of the dielectric circular cavity with point scatterer $(n=3,d=0.59)$ for
$\Real k \in [0,4]$. The solid (red) lines are the paths of the perturbed resonances as the strength
varies between $a=0$ and $a=\infty$. The dots are the unperturbed resonances of the cavity without scatterer
and the crosses are the perturbed resonances for $a=0.1$.  For $a=0.1$, the position of the first perturbed resonance is not visible as its location is deep
in the complex plane at $k=9.5670\times10^{-7} - 3.7435 \ui$.
}
\end{figure}

As we vary $a$ the perturbed resonances follow paths in the complex plane, each of which connects together either two
different unperturbed resonances or one unperturbed resonance to itself. An example is given in Fig.~\ref{fig:leveldyn_lowk}
for low values of $k$. We also show in the plot the unperturbed resonances of Section \ref{sec:unptbd_open_exact} and the
perturbed resonances corresponding to a fixed perturbation $a=0.1$. We see the first perturbed resonance is moving towards
the imaginary axis and then much deeper into the complex plane. Some of the paths in Fig.~\ref{fig:leveldyn_lowk} connect
individual unperturbed resonances to themselves and the others connect different unperturbed resonances. There are no
instances of a pair of unperturbed resonances being connected in a loop-like manner. None of the paths cross (due to level
repulsion) but there can be instances of two paths becoming very close to each other, and hence a fine discretisation of $a$
is required. Furthermore, the choice of the location of the scatterer $d$ determines the overall structure of the connections
\cite{Dettmann2009uni}.

Because many of the unperturbed resonances had their imaginary parts set to zero in the previous section, the computation of
the corresponding perturbed resonances is complicated as the quantisation condition (\ref{eq:quantcond_pted}) is technically
unsatisfied at the start of the path. Moreover, paths that start at one of these unperturbed resonances near the real axis do
not necessarily stay in the vicinity of it and hence we cannot arbitrarily take the perturbed resonance to be equal to the
unperturbed resonance. We verify this by introducing the set $\sigma_{\epsilon}$ of unperturbed resonances $k = k_r + \ui k_i$
with $-k_i < \epsilon$. We then omit this set of resonances from the computation, vary $a$ from 0 to $\infty$, which is numerically
taken to be $a=10^{30}$, and identify the presence of any unperturbed resonances $k \in \sigma \backslash \sigma_{\epsilon}$
(that lie deeper in the complex plane) to which no level path has reached. The singular nature of the condition (\ref{eq:quantcond_pted})
at the unperturbed resonances prohibits a perturbed resonance from numerically reaching those values. We hence determine the
`equality' of two such resonances by using the approximate solution (\ref{eq:small_perturb}).

We call this set of unperturbed resonances $\sigma'$ and on each element within it perform the perturbation computation backwards
from $a=\infty$ to $a=0$ where we always end up at the real axis. The unperturbed resonance in $\sigma_{\epsilon}$ which lies
closest (up to using (\ref{eq:small_perturb})) to each termination point of each point on the real axis is then removed from
$\sigma_{\epsilon}$. The resulting set $\sigma_{\epsilon}$ then contains only unperturbed resonances which, when perturbed,
follow tiny loops near the real axis back to themselves and is hence the set for which we take the perturbed resonance to be
equal to the unperturbed resonance, regardless of the value of $a$. The perturbation computation is then repeated for
$a=0 \rightarrow a_{0}$ on the set $\sigma \backslash \sigma_{\epsilon}$ and for $a = \infty \rightarrow a_{0}$ on the set $\sigma'$,
the results of which contribute, along with the set $\sigma_{\epsilon}$, to the desired set of all perturbed resonances $\tilde{\sigma}$
in (\ref{eq:ptbd_res_open}).

\begin{figure}
\centerline{
\includegraphics[angle=0,width=12cm]{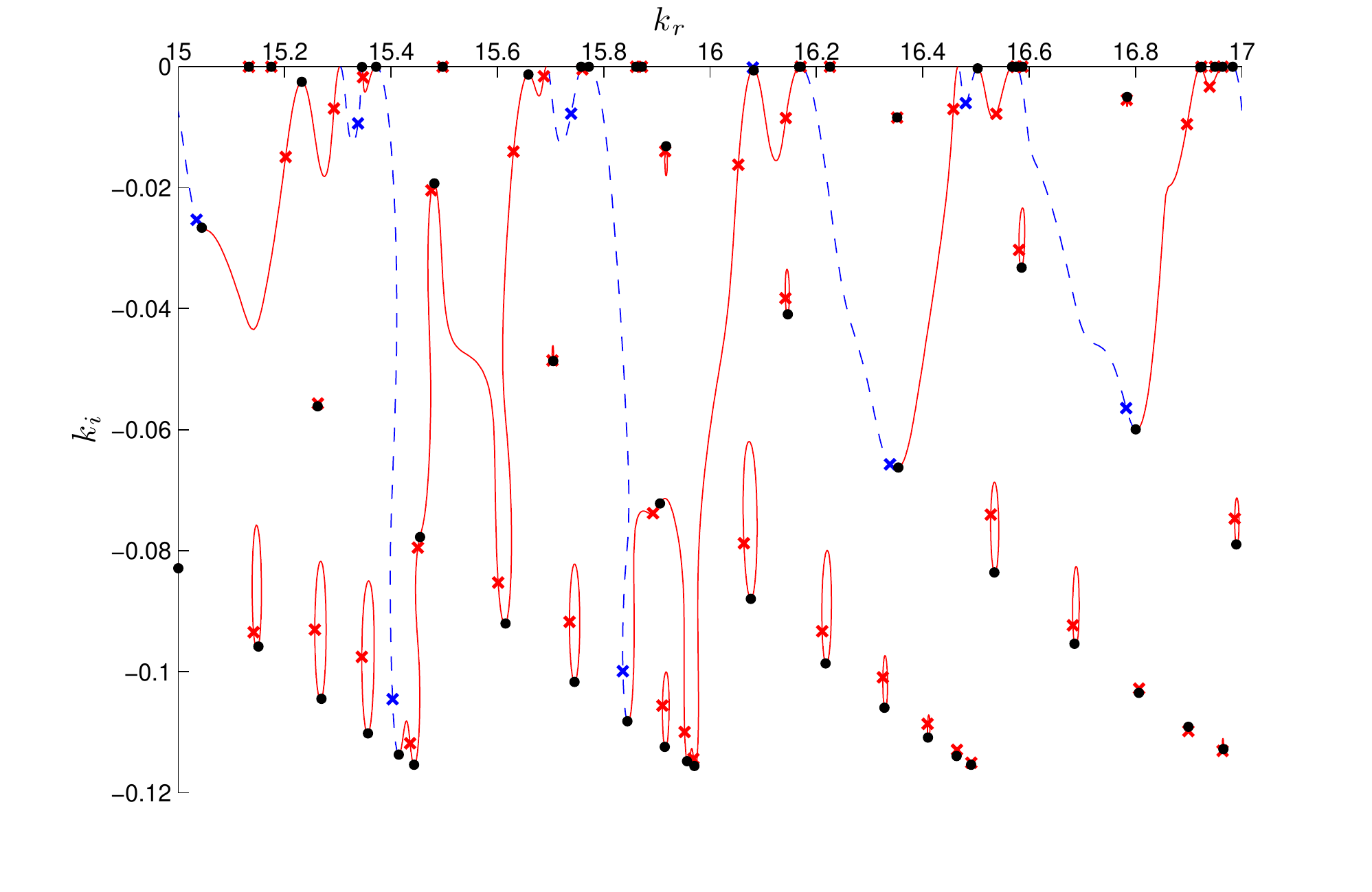}
}
\caption[Level dynamics of the perturbed resonances for $\Real k=15...17$.]{\label{fig:leveldyn_highk}
Level dynamics of the perturbed resonances of the dielectric circular cavity with point scatterer $(n=3,d=0.59)$ for
$\Real \, k \in [15,17]$. The solid (red) lines are the paths of the perturbed resonances in the set
$\sigma \backslash \sigma_{\epsilon}$ as the strength varies from $a=0$ to $a=\infty$ and the dashed (blue)
lines are the level dynamics of the resonances in the set $\sigma'$ which travel to unperturbed resonances
having small imaginary part as the strength varies from $a=\infty$ to $a=0$ (see text). The dots are the
unperturbed resonances of the cavity without scatterer and the crosses are the perturbed resonances for $a=0.1$.
}
\end{figure}

In Fig.~\ref{fig:leveldyn_highk} we show the level dynamics of perturbed resonances for $k$ in the range 15 to 17, and illustrate
the above method by highlighting level curves and resonances as computed from the sets $\sigma \backslash \sigma_{\epsilon}$
or $\sigma'$ separately. As an illustration, for the calculations in the next section we will use resonances in the range
$\Real \, n k \in [0,100]$ for $n = 3$, $d = 0.59$, and $a = 0.1$. In our calculations to find them we used 1278 unperturbed
resonances (of which 32 had $m=0$), the set $\sigma_{\epsilon}$ contained 432  perturbed resonances, the set
$\sigma \backslash \sigma_{\epsilon}$ contained 750 and the set $\sigma'$ contained 98, with a total of 1280.
The discrepancy of 2 comes from an extra perturbed resonance at each end of the wavenumber interval.

Because the diffraction coefficient entering the semiclassical amplitudes changes with the additional factor of $n$, the
optimal value of the strength of the scatter $a$ in a given $k$-range also changes. We found that taking $a=0.1$ gives
strong diffractive peaks in the length spectra, which is smaller than the value $a=1$ taken for the corresponding closed
system. In our plots we use the same scatterer position $d=0.59$ as in the closed billiard.

\begin{figure*}
\centerline{
\includegraphics[angle=0,width=10cm]{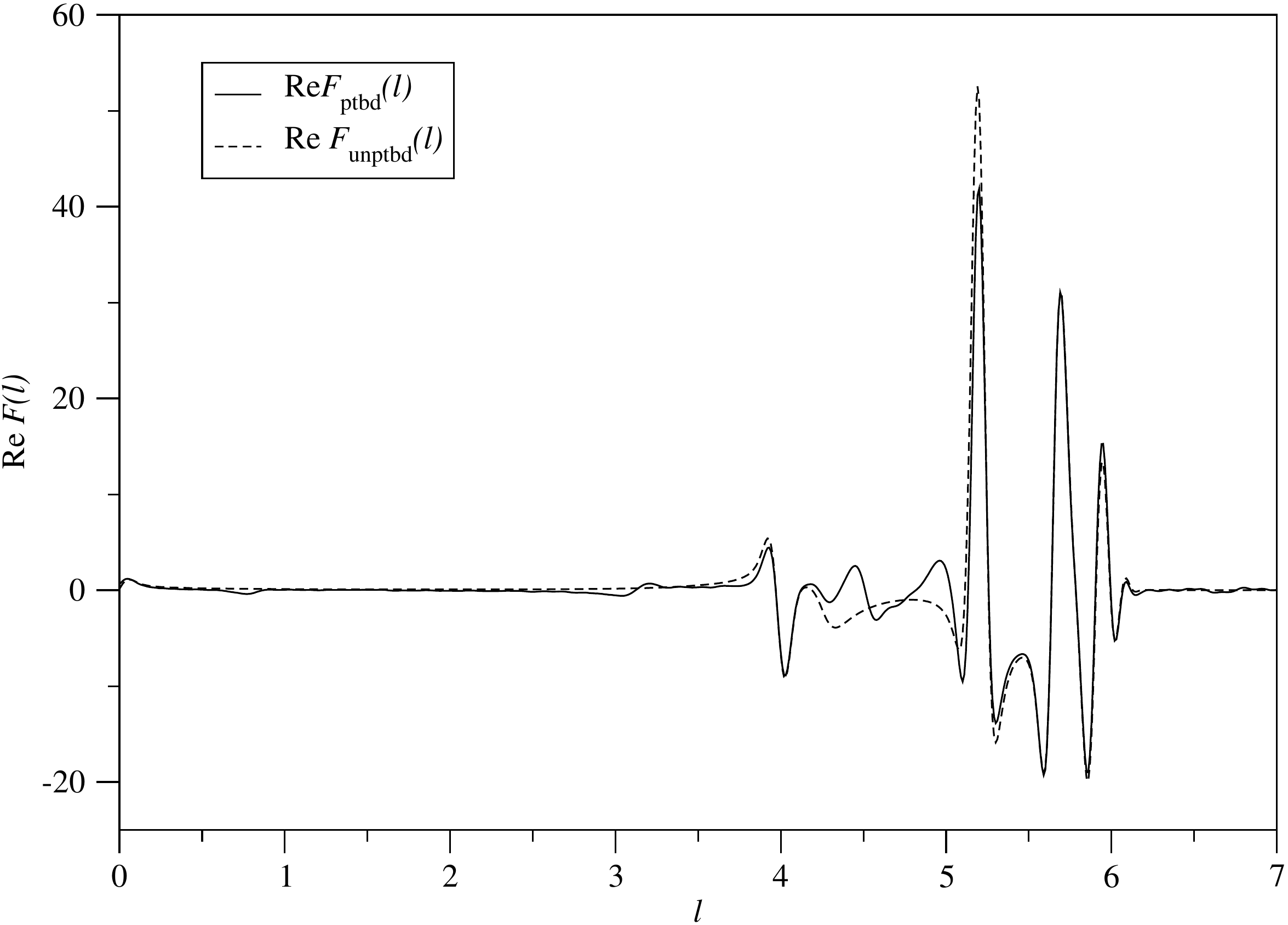}
}
\caption[Unperturbed and perturbed exact length spectra.]{\label{fig:length_spec_unptbd_ptbd_open}
Exact length spectra for the unperturbed (full line) and perturbed (dashed line) circular dielectric cavities.
($d=0.59$, $a=0.1$, $t=0.001$, $\mu_{\mathrm{max}} = 0$.) 
}
\end{figure*}

The exact length spectrum is of the same form as before and is given by
\begin{equation}\label{eq:trace_qm_fourier_open_ptbed}
\tilde{F}(l) = \int_0^\infty k^{\mu_{\mathrm{max}/2}} \tilde{d}(k) \, W(k) \expp^{- \ui n k l} \ud k \,,
\end{equation}
where $\tilde{d}$ is the level density of the perturbed system. 
In Fig.~\ref{fig:length_spec_unptbd_ptbd_open} we plot the perturbed length spectrum (\ref{eq:trace_qm_fourier_open_ptbed})
alongside the unperturbed spectrum (\ref{eq:trace_qm_fourier}) for a scatterer position $d=0.59$, scatterer strength $a=0.1$
and $\mu_{\mathrm{max}}=0$. As will be discussed in the next section the difference can be attributed to diffractive orbits.
In order to visualize the signature of diffractive orbits in the length spectrum we will again consider in the following the difference, $\Delta F(l) = \tilde{F}(l) - F(l)$, between perturbed and unperturbed length spectra .

\subsection{Short-wave approximation} \label{sec:ptbd_open_sc}

Similar to the modification of the periodic orbit terms in the trace formula when we moved from the closed to the open system,
we must also modify the diffractive orbit terms. According to equation (\ref{eq:d_xi2}) this consists of adding an overall factor
of $n$ to the diffractive part (\ref{eq:diff_conbution}) of the closed system, changing $k$ to $n k$, and replacing the Dirichlet
phase factors by Fresnel reflection coefficients. This results in

\begin{equation} \label{eq:diff_conbution_open}
d_{\mathrm{do}}(k) = \sum_d A_d \exp \left\{ \ui \left(n k L_d - \frac{\nu_d \pi}{2} - \frac{3 \mu_d \pi}{4} \right) \right\}
+ \mathrm{c.c.} \,,
\end{equation}
where the amplitudes are given by
\begin{equation} \label{eq:amplitude_open}
A_d = \frac{n l_d}{2 \pi} \left \{\prod_{j=1}^{\mu_d} \frac{g_{d,j} \mathcal{D}(n k)[R(\theta_{d,j})]^{r_{d,j}} }{
\sqrt{8 \pi n k \vert(M_{d,j})_{12} \vert}} \right \} \,,
\end{equation}
and the sum runs over all diffractive orbits as presented in Section \ref{sec:ptbd_closed_sc}. The refection coefficient is
now evaluated at the angle of incidence of each contributing diffractive orbit (which is complex valued for ghost orbits
in a tangent bifurcation). The diffraction and reflection coefficients are both complex valued.

The difference $\Delta F(l)$ between the length spectra of perturbed and unperturbed systems is given only in terms
of the diffractive orbits in the short-wave approximation, because the periodic orbit terms are the same in both systems
and cancel. Hence we obtain the approximation
\begin{equation} \label{eq:trace_sc_fourier_open_diff}
\Delta F(l) \approx \int_0^\infty k^{\mu_{\mathrm{max}/2}} d_{\mathrm{do}}(k) W(k) \expp^{-\ui n k l} \ud k \,,
\end{equation}
which has to be evaluated numerically. The quantity $\mu_{\mathrm{max}}$ is again the chosen maximum number of
encounters with the scatterer of the diffractive orbits that we enter into the sum. In (\ref{eq:trace_sc_fourier_open_diff})
we neglect the small modification of the smooth part of the density of states due to the scatterer.

\begin{figure}
\centerline{
\includegraphics[angle=0,width=12cm]{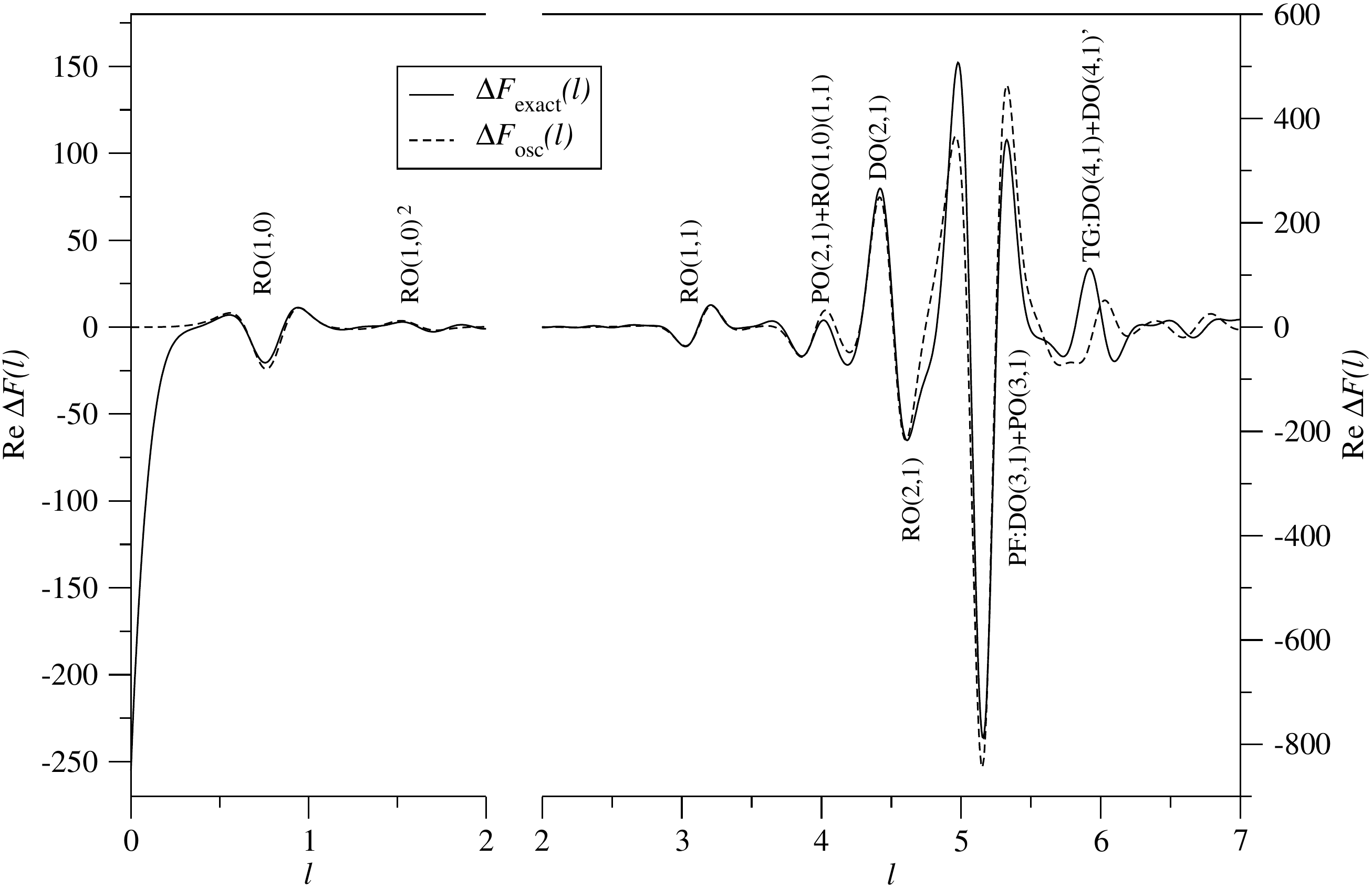}
}
\caption[Exact and semiclassical length spectrum differences.]{\label{fig:length_spec_diff_open}
Difference between length spectra for perturbed and unperturbed dielectric cavities. The exact result (full line)
is compared to the short-wave approximation (dashed line) that is obtained from diffractive orbits
using uniform approximations. ($d=0.59$, $a=0.1$, $\mu_{\mathrm{max}}=4$, $t=0.005$.)
}
\end{figure}

In Fig.~\ref{fig:length_spec_diff_open} we plot the length spectrum difference
alongside its semiclassical counterpart (\ref{eq:trace_sc_fourier_open_diff}) given above. We see that the peaks
in $\Delta F(l)$ can clearly be identified with contributions of the diffractive orbits. We found it necessary to choose a higher
scaling factor $t=50 / k_{\mathrm{max}}^2$ than in previous sections, because we encountered larger fluctuations.

As with the perturbed circular billiard, the diffractive contribution (\ref{eq:diff_conbution_open}) to the trace formula
breaks down near bifurcations of the diffractive orbits, and we again applied uniform approximations to deal with
these cases. The bifurcation scenarios are the same as for the circular billiard and are given in
Tab.~\ref{tab:bifurcations} and we use the same uniform approximations as for the closed system, modifying the
amplitudes of the orbits involved in each bifurcation appropriately. The semiclassical curve in Fig.~\ref{fig:length_spec_diff_open}
includes these uniform approximations. We have again included the uniform approximation for the tangent bifurcation
of the $\mathrm{DO}(4,1)$ and $\mathrm{DO}(4,1)'$ orbits, and for the pitchfork bifurcation involving the orbits
$\mathrm{DO}(3,1)$ and $\mathrm{PO}(3,1)$. We see in the figure that there is a greater discrepancy
between the exact and semiclassical length spectra corresponding to these orbits than in the analogous figure
(Fig.~\ref{fig:length_spec_diff_closed}) for the closed billiard.

In the previous section we saw that the semiclassical approximation to the length spectrum of the unperturbed
dielectric cavity is poorer at lengths corresponding to periodic orbit whose angle of incidence with the boundary is near the
critical angle. We gave a formula (\ref{eq:po_crit_angle}) that specifies this length as a function of the refractive
index $n$. We now perform a similar calculation for diffractive orbits. If we insert the formula for the critical angle
into the formula for the diffractive orbit lengths (\ref{eq:diff_orb_length}), restrict to $w=1$ rotations and use an
interpolating formula for the number of reflections $r$ in terms of the angle of incidence we obtain
\begin{equation} \label{eq:do_crit_angle}
l_d(n) = \frac{2}{n} \left[ \left( \frac{2 \arcsin(1/n d) + \pi}{\pi - 2 \theta_c(n)} \right) \sqrt{n^2 - 1} + \sqrt{d^2 n^2 - 1} \right] \,,
\end{equation}
which is defined for $n \geq 1/d$. At the limiting value $n=1/d$ it agrees with $l_p(n)$ in (\ref{eq:po_crit_angle}), and
$l_d(n) \rightarrow 2(1+d)$ as $n \rightarrow \infty$. For our numerics, the relevant value is $l_d(3) \approx 4.30$
for $d=0.59$. At this length we would expect an inaccuracy in the diffractive contribution to the semiclassical
length spectrum.

\begin{figure}
\centerline{
\includegraphics[angle=0,width=10cm]{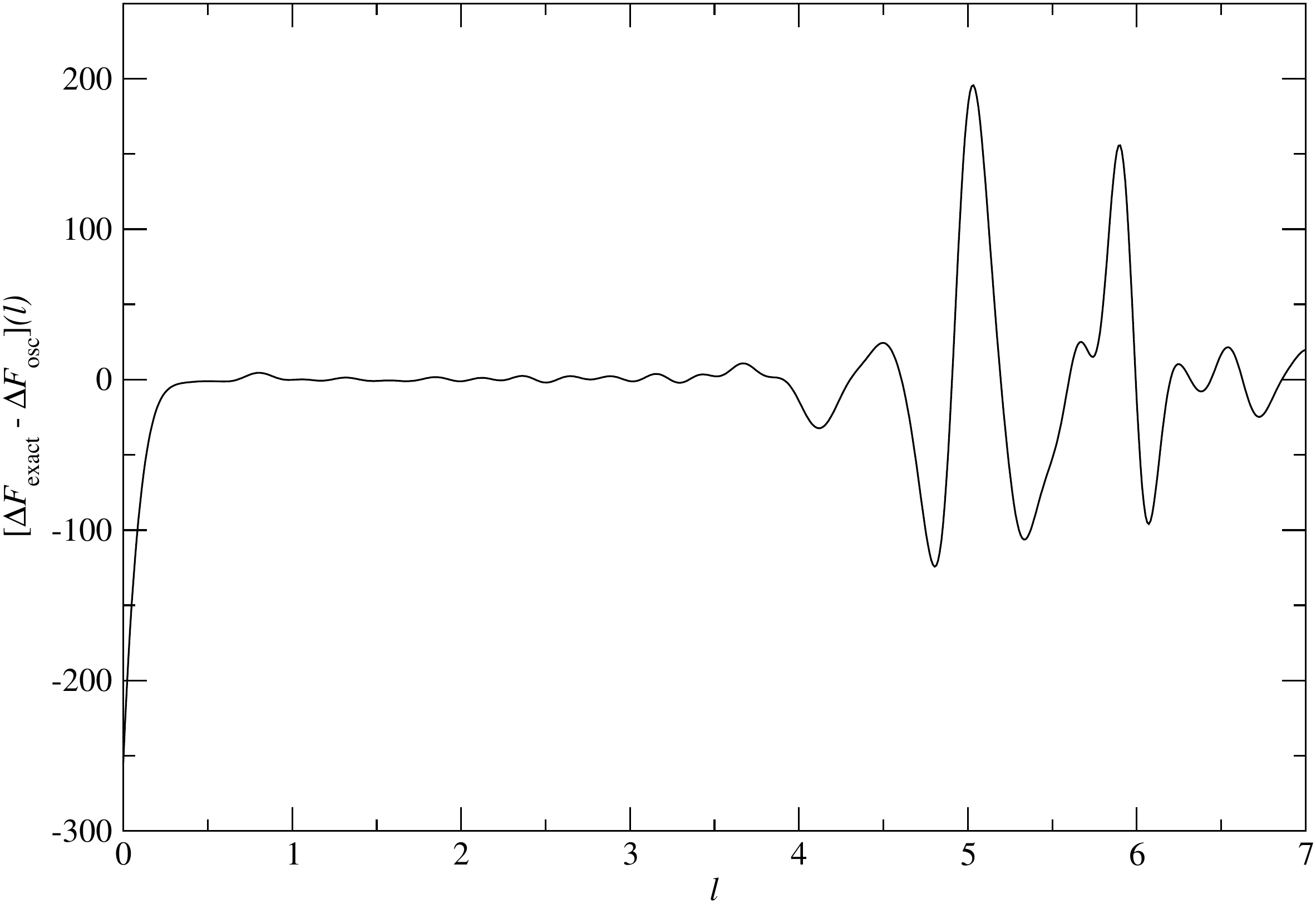}
}
\caption[Discrepancy between the exact and semiclassical length spectrum differences.]{\label{fig:length_spec_diff_diff}
Discrepancy between the exact and semiclassical length spectrum differences in Fig.~\ref{fig:length_spec_diff_open}
($d=0.59$, $a=0.1$, $\mu_{\mathrm{max}}=4$, $t=0.005$.)
}
\end{figure}

In Fig.~\ref{fig:length_spec_diff_diff} we show the discrepancy between the exact and semiclassical length spectra differences
of Fig.~\ref{fig:length_spec_diff_open}. The error at the length that corresponds to the critical angle does not play such a big role
in this plot. The discrepancy comes mainly from the fact that the peak heights are not completely reproduced by the semiclassical
plot, as can be seen in Fig.~\ref{fig:length_spec_diff_open}. As discussed before, however, there is also a slight uncertainty about
the numerical error of some of the resonances near the real line. All in all, the semiclassical approximation captures the peak
structure of the length spectrum quite well.

\section{Conclusions} \label{sec:conc}

The aim of this article was an investigation of trace formulas for dielectric cavities. 
Such a trace formula was proposed recently for TM modes in quasi two-dimensional
geometries, and it connects the spectrum of the Feshbach or interior resonances 
in the short-wave limit to the ray dynamics inside the cavity \cite{Bogomolny2008}.

For the dielectric circle we evaluated the Fourier transform of the resonance 
spectrum, the length spectrum, which has peaks at the periodic rays inside
the cavity. With more than 2000 resonances we could probe the short-wave
regime. In contrast to previous studies we compared not only the position
but also the shape of the peaks. We found that the agreement with the
short-wave approximation is remarkably good, except for expected inaccuracies
for orbits with reflection angles near the critical angle for total internal reflection.
Otherwise, the approximation is of similar quality as for the closed cavity.
We also saw good agreement for orbits which are not in the regime of total
internal reflection which were found difficult to see in previous articles
\cite{Bittner2010,Bogomolny2010}.

We then considered a dielectric circular disk with an additional point scatterer.
This required a generalization of the trace formula which includes additional
contributions from diffractive rays that start and end at the scatterer.
This system allows for a more probing test of the trace formula. On the
one hand it is a non-integrable system, but one can nevertheless
determine a large number of resonances by a Green function method.
On the other hand the contributions of isolated diffractive rays are at
least an order of $1/k$ smaller than those of the continuous families
of periodic orbits in the circle. For example, in \cite{Bogomolny2010}
it was found difficult to get a good quantitative agreement with the trace
formula in systems with isolated periodic orbits. 

In order to isolate the peaks at diffractive orbits from those at periodic orbits
we considered the difference of the length spectra for dielectric circular disks with and
without scatterer. The short-wave approximation for this difference is solely
given by diffractive orbits. Our numerical results showed that the short-wave
approximation works well also in this case. The position and shape of
the peaks are well reproduced by the trace formula. Only the agreement
in the height of the peaks was not as good as in the comparable case
of the closed cavity with a point scatterer. Some diffractive orbits were
close to a bifurcations and we improved the approximation by using
uniform approximations. We also saw clear peaks at diffractive orbits
which are not in the regime of total internal reflection. 

In summary, we found that trace formulas for dielectric cavities work
well in the short-wave regime. Even higher order contributions from
diffractive rays can clearly be identified. One possible improvement
is to include modifications of the Fresnel coefficient
for reflections near the critical angle \cite{Bogomolny2010}. As a 
possible application of our results one could use the resonance
spectrum to locate the position of a defect within an optical cavity.

%\bibliographystyle{alpha}
%\bibliography{microcavity_paper_bib}

\newcommand{\etalchar}[1]{$^{#1}$}

\end{document}